\documentclass[aip,reprint]{revtex4-1}

\usepackage{amsmath}    
\usepackage{graphicx}   
\usepackage{verbatim}   
\usepackage{color}      
\usepackage{hyperref}   
\usepackage[utf8]{inputenc} 

\begin{document}

\title{Towards dynamical network biomarkers in neuro\-modulation of episodic migraine}

\author{Markus A.\ Dahlem}
\email[]{dahlem@physik.hu-berlin.de}
\author{Sebastian Rode}
\affiliation{Department of Physics, AG NLD Cardiovascular Physics, Humboldt-Universit\"at zu Berlin, Robert-Koch-Platz 4, 10115 Berlin, Germany}

\author{Arne May}
\affiliation{Center for Experimental Medicine,  Department of Systems Neuroscience, Universit\"atsklinikum Hamburg-Eppendorf, 20246 Hamburg, Germany}

\author{Naoya Fujiwara}
\affiliation{FIRST, Aihara Innovative Mathematical Modelling Project, Japan Science and Technology Agency}

\author{Yoshito Hirata}
\author{Kazuyuki Aihara}

\affiliation{Collaborative Research Center for Innovative Mathematical Modelling, Institute of Industrial Science, University of Tokyo, Tokyo 153-8505, Japan}

\author{J\"urgen Kurths}
\affiliation{Department of Physics, AG NLD Cardiovascular Physics, Humboldt-Universit\"at zu Berlin, Robert-Koch-Platz 4, 10115 Berlin, Germany}
\affiliation{Potsdam Institute for Climate Impact Research, 14473 Potsdam, Germany}
\affiliation{Institute for Complex Systems and Mathematical Biology, University of Aberdeen, Aberdeen AB24 3UE, United Kingdom}

\date{\today}

\begin{abstract} 
Computational methods have complemented experimental and clinical neursciences
and led to improvements in our understanding of the nervous systems in health
and disease.  In parallel, neuromodulation in form of electric and magnetic
stimulation is gaining increasing acceptance in chronic and intractable
diseases. In this paper, we firstly explore the relevant state of the art in
fusion of both developments towards translational computational neuroscience.
Then, we propose a strategy to employ the new theoretical concept of dynamical
network biomarkers (DNB) in episodic manifestations of chronic disorders. In
particular, as a first example, we introduce the use of computational models in
migraine and illustrate on the basis of this example the potential of DNB as
early-warning signals for neuromodulation in episodic migraine.
\end{abstract}

\maketitle
 
\section{From bifurcation to bench and bedside}
\label{sec:b2b}

The mathematical theory of bifurcations is part of dynamical systems theory,
which can straightforwardly be applied to neurological and psychiatric diseases
with episodic manifestations.  The term `bifurcation' in this context refers to
abrupt changes of brain dynamics when some physiological parameter values are
smoothly changed and, at a critical point---the bifurcation point---the
qualitative structure of the dynamical behavior changes towards these new
pathological dynamics.  When physiological control is lost in such a scenario,
the pathological process is called a dynamical disease\cite{MAC87}. Epilepsy is
the best studied neurological dynamical disease where the theoretical concepts
of bifurcation theory have been translated to animal models (bench) and
clinical studies (bedside)\cite{MIL03a}.

The theory of such  bifurcations, or ``tipping points''\cite{SCH09g},
complements experimental and clinical neurosciences with mathematical analysis
and simulations to interpret data and guide a principle understanding of the
nervous systems in health and disease.  One particular interest is in defining
dynamical network biomarkers (DNB) as early-warning signals based on an
analytical understanding of the behavior at the imminent tipping
point\cite{CHE12a,LIU12b,LIU13a,LIU13b}. We will focus on this in this paper
and firstly extend this concept to obtain a framework for a systematic
evaluation of the subtle symptoms that often occur in the prodromal phase
before the main episodic manifestations of chronic disorders.  In a second
step, we suggest to use DNB beyond monitoring early-warning signals and explore
ways in which DNB can lead to  biofeedback signals in a control paradigm for
episodic treatment.

\begin{figure*}[t] \begin{center}
\fbox{\includegraphics[width=\textwidth]{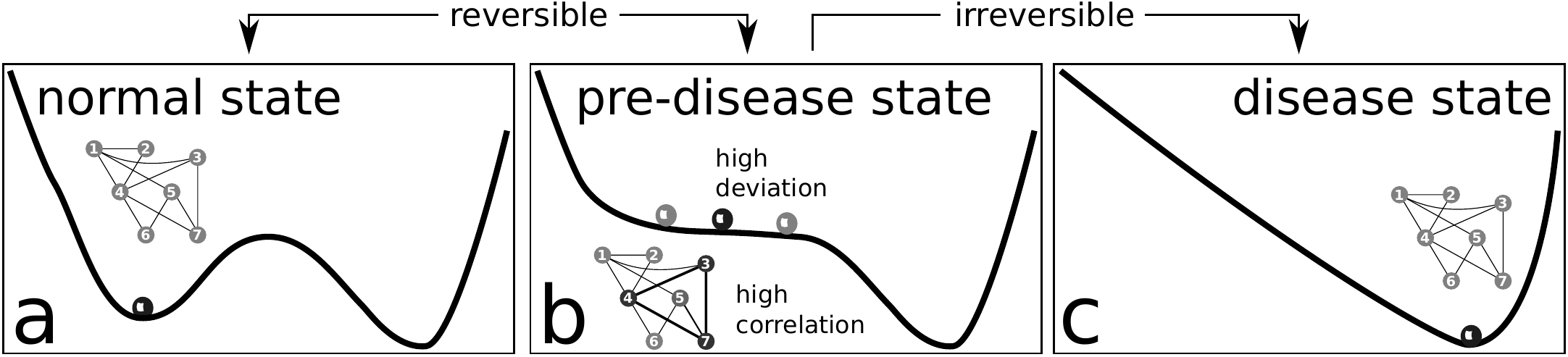}} \end{center}
\caption{\label{fig:dnb} Schematic illustration of the dynamical features of
disease progression from a normal state to a disease state through a
pre-disease state (modified from Ref.~\cite{CHE12a}). (a) The normal state is a
steady state or a minimum of a potential function, representing a relatively
healthy stage.  (b) The pre-disease state is situated immediately before the
`tipping point'. At this stage, the system is sensitive to external stimuli and
still reversible to the normal state when appropriately interfered with. A
small change in the parameters of the system may suffice to drive the system
into collapse, which often implies a large phase transition to the disease
state. When the system approaches the pre-disease state, the deviations of the
set of nodes (3, 4, 7) increase drastically, and the correlations (positively
or negatively) among them also increase drastically whereas their correlations
with other nodes (1, 2, 5, 6) decrease drastically.  We call (3, 4, 7) the DNB.
(c) The disease state is the other stable state or a minimum of the potential
function, where the disease has seriously deteriorated and thus the system is
usually irreversible to the normal state.  } \end{figure*}

One central aim in translational computational neuroscience is to fuse
dynamical systems theory with control theory\cite{AIH10} and drive innovation
in therapeutic brain stimulation in neurological and psychiatric diseases with
theoretical concepts. At this stage, let us only name two further dynamical
diseases other than epilepsy\cite{MIL03a}, firstly (i), Parkinson's disease,
which we do not further explore but see Ref.~\cite{SCH10h} and, secondly (ii),
primary headaches, which we set the focus on in this paper. In these dynamical
diseases, theoretical concepts are used (i) to illuminate the role of the
coupling architecture and abnormal activity patterns in the subthalamopallidal
network of the basal ganglia  and with that knowledge to explain the mechanisms
underlying the therapeutic efficacy of deep brain stimulation in Parkinson's
disease \cite{TER02,RUB12}, and (ii) to understand the dynamics of pain as a
process of central sensitisation in the migraine generator network and define
neuromodulation targets and stimulation protocols for episodic migraine
treatment \cite{DAH13} that possibly can also be explored for neuromodulation
in cluster headache \cite{MAG12}, though we will focus mostly  on migraines. 

A brief note on the history is due. Despite the fact that the fields of
translational computational neuroscience and neuromodulation are still in their
relative infancy, let alone their fusion, they both actually have a long partly
overlapping history.  Neuromodulation for example in form of non-drug treatment
in headache dates back to the mid-first century CE to Scribonius Largus, a
physician who practiced at the court of the Roman Emperor Claudius and who
recognized the seemingly beneficial effects of discharges from electrical fish
in headache\cite{KOE10}. Not astonishingly, these primitive neuromodulation
concepts gained much new interest in the 19th century, when Emil du Bois
Reymond---himself a migraine sufferer---devised the first very precise
instruments for measuring and monitoring electrical biosignals and  his pupil
Julius Bernstein developed the first membrane theory of electrical resting and
action potentials in neurons.  At the same time, driven by the industrial
revolution James Clark Maxwell invented closed-loop feedback to control the
speed of steam engines, which provides the first theoretical framework that
later led to the field of cybernetics as the scientific study of ``control and
communication in the animal and the machine''\cite{WIE48}.  The modern aspects
of this field with some further notes on the history have been recently
described in the textbook entitled ``Neural Control Engineering: The Emerging
Intersection Between Control Theory and Neuroscience''\cite{SCH11e}.\\

In the following, we first briefly introduce dynamical network biomarkers
(Sec.~\ref{sec:dnb}) and then adapt the concept DNB to chronic disorders with
episodic manifestations (Sec.~\ref{sec:dnbcdem}) and highlight migraine as an
example (Sec.~\ref{sec:phae}). To this end, we introduce migraine theories
(Sec.~\ref{sec:mt}) and computational models of spreading depression
(Sec.~\ref{sec:cm}) on microscopic and macroscopic scale where the concept of
DNB must be applied. We end with an overview of neuromodulation and concluding
remarks (Sec.~\ref{sec:nm}).

\section{Dynamical network biomarkers} \label{sec:dnb}

Traditional biomarkers are some objectively measured and evaluated indicators
of a particular biological state that occurs in association with a pathological
process. Usually static measurements of some traceable substances are called
biomarkers. Also genetic biomarkers exist, defined as mutations or
polymorphisms that predict some clinically relevant measure for example  risk
of disease, its  outcome or response rate to treatment.

In contrast, a DNB is a dynamical feature of a biological network under
consideration.  Such features  were originally suggested to be of particular
interest in complex diseases with sudden deterioration phases or critical
transition points during their progressions\cite{CHE12a,LIU12b}.  Depending on
the progression level of such disorders three stages are distinguished: a
normal state, a critical pre-disease state, and a disease state (Fig. 1). The
normal state is a steady state with functional homeostatic control,
representing a (relatively) healthy stage yet maybe in an incubation period.
The critical pre-disease state is still clinically silent and is considered as
the state just before the first clinical onset.  Mathematically, the critical
pre-disease state is defined as the limit of the normal state immediately
before a well defined tipping point is reached.  Finally, the disease state is
with control failure but otherwise, in particular from a mathematical
perspective, also simply a stable state. 

A DNB is essentially a group of traceable substances or, in general, signals
that while highly fluctuating are strongly correlated only during the
pre-disease stage.  Therefore a DNB is different from the conventional
biomarkers in two ways. First, it is not required for DNBs to keep consistent
values for the respective disease and normal samples. Only the presence of
highly fluctuating and strongly correlated signals is important. Second, the
aim of DNB is to detect the pre-disease state, i.e., early-warning signals
associated with the imminent tipping point, but not to distinguish the
respective disease and normal states.

If the pre-disease state is detected by a DNB, early treatment can be started,
see Fig.~\ref{fig:dnb}. It was first analytically shown that a DNB serves as an
early-warning signal by defining particular but rather abstract critical states
as coherent subnetworks that were then recognized as the physiological
correlate of the pre-disease state. DNBs were successfully identified in
medicine for lung injury disease, liver cancer, and lymphoma
cancer\cite{CHE12a,LIU12b}.

\section{Chronic disorders with episodic manifestations --- dynamical systems
theory view} \label{sec:dnbcdem}

\begin{figure}[t] \begin{center}
\fbox{\includegraphics[width=\columnwidth]{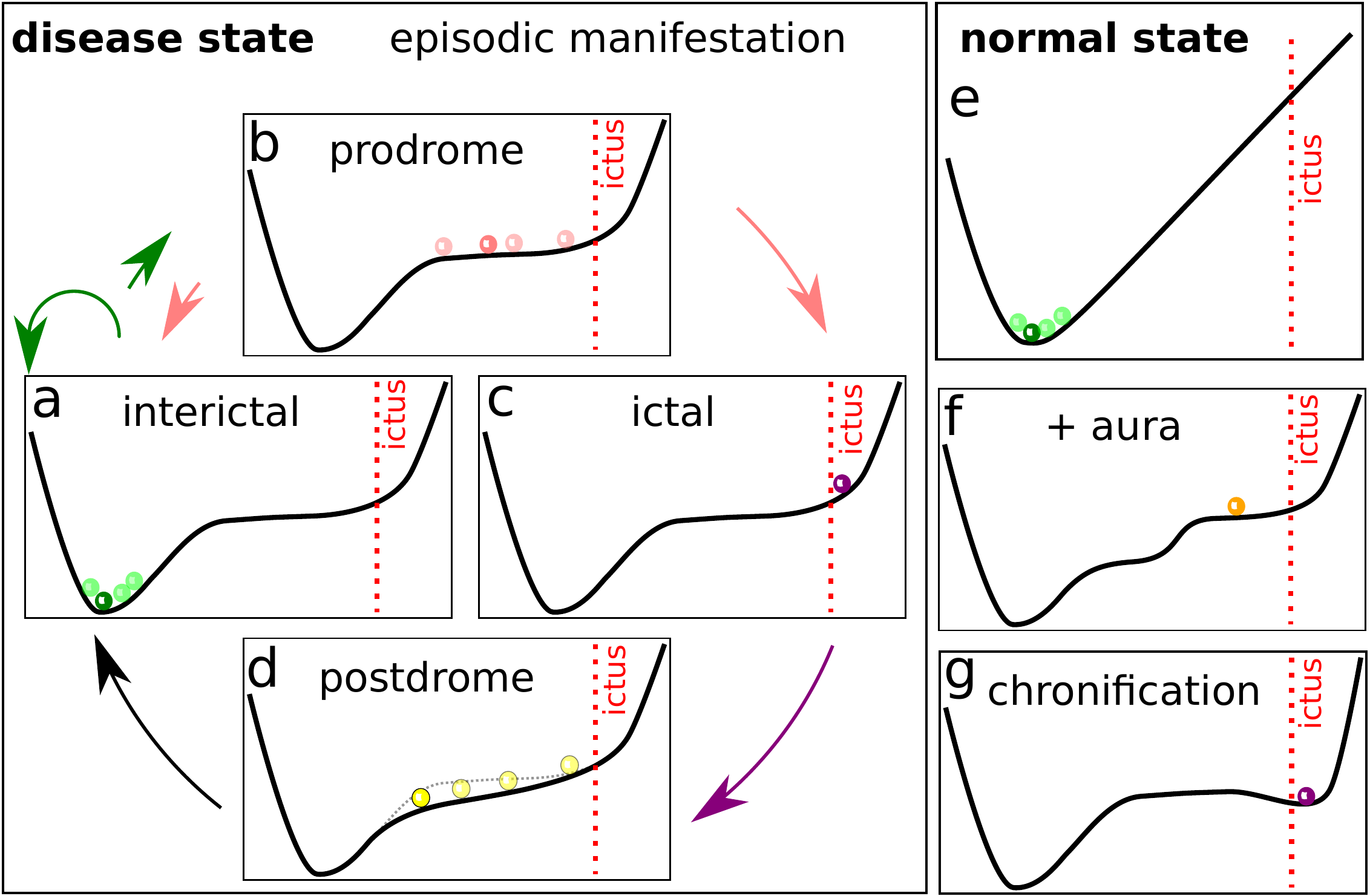}}
\end{center} \caption{\label{fig:cdem-mechanism} Possibly dynamical scenario of
recurrent episodes as a cycle (a)-(d) in chronic disorders (disease state) vs
normal state (e) described in pictorial `landscapes'. (a) interictal: stable
state where the disease under control; (b) prodrome: when the state escapes
over the potential barrier to a relative flat plateau, the transient meta
stable state is characterised by a strongly fluctuating and correlated
subnetwork (cf.~Fig.~\ref{fig:dnb}); (c) ictal: at a certain point, the
episodes manifest in form of a transient excursion that is recovered in (d)
postdrome (possibly changed landscape); (e) normal state remains stable because
the potential barrier toward th ictal state is too high. (f) Additional
premonitory symptoms such as migraine aura or aura in epilepsy indicates the
existence of different plateaus (cf.~Fig.~\ref{fig:combinedMechanism}). (g)
Chronification of diseased state: landscape changes such that the attack state
persists.} \end{figure}

The idea of a transition in form of a sudden deterioration from the health to
the disease state via a specific pre-disease state can be transfered to the
cycle of recurrent episodic symptoms in chronic disorders.  The pre-disease
state corresponds there to the prodrome, a stage that is relatively clinically
silent but subtle symptoms announcing the ictal phase. It is worthwhile noting
that the theoretical concepts are completely independent of the application and
can be transfered to various complex systems. Early-warning signals near
tipping points exist in systems ranging from medicine to financial market (and
their crashes), to power grid systems to which a large amount of renewable
energy is introduced, to ecosystem, and to the global climate system
\cite{SCH09g}.  

Many common neurological and psychiatric disorders fit into the category of
chronic disorders with episodic manifestations including not only migraine and
epilepsy but also stroke, multiple sclerosis, sleep-wake disorders, addictive
disorders, schizophrenia and depression. But, it remains to be investigated
which are likely candidates for dynamical diseases, which, we define here,
accompany abrupt changes in natural rhythms in some organ systems due to
bifurcations.

The existence of a prodrome with subtle symptoms before the ictal phase is a
promising indicator of such a mechanism based on a bifurcation that shows
early-warning signals. A prodrome stage is actually what would be expected from
DNB being a subnetwork that is both highly fluctuating and strongly correlated
but still in the physiological range.   So let us consider again the two ways
described before in which DNB are different from conventional biomarkers. If the
concept is transfered to chronic disorders with episodic manifestations, DNB
are by definition only present during the prodrome. Therefore, the DNB should
be considered as the neural correlate of the prodrome. In other words, the
existence of DNB is not a prediction in these cases but rather one possible
{\em a posteriori} explanation for the prodrome.

\begin{figure}[t] \begin{center}
\fbox{\includegraphics[width=\columnwidth]{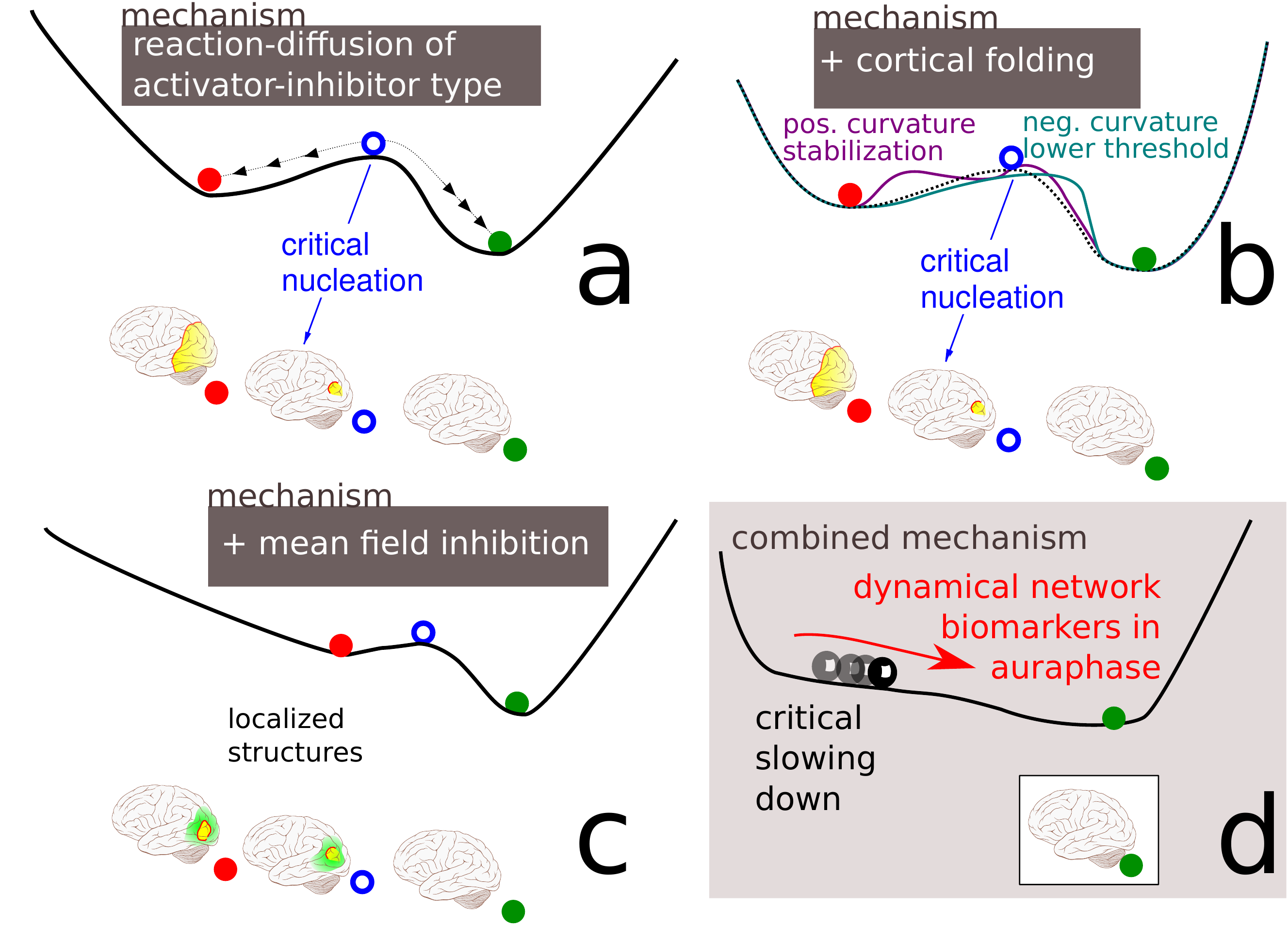}}
\end{center} \caption{\label{fig:combinedMechanism} Schematic illustration of
the `landscape' in a macroscopic model of SD (for details see
Sec.~\ref{sec:tissue}). (a) Reaction-diffusion systems describe excitable media
that support traveling waves and can engulf a cortical hemisphere. (b) Cortical
folding modulates the pattern forming features. (c) Additional mean field
control can stabilize localized SD structures. (d) A critical slowing down of
localized SD structures is observed in a model with large mean field feedback
control and the patterns are modulated (breathing spots) in curved geometries.}
\end{figure}

The cycle of recurrent episodic manifestations is illustrated in a simplified
way in Fig.~\ref{fig:cdem-mechanism}. The states are associated to a potential
function with a steady state or a minimum of this potential function
corresponding to the attack-free interval and a prodrome phase corresponding to
a plateau level, which can be reached by spontaneous fluctuations that are able
to escape the potential barrier from the minimum level to the plateau level. At
the far end of the  plateau level, the `ictus' is marked as the state that
initiates the attack.  The plateau leads to long transients and DNB dynamics,
see also Fig.~\ref{fig:combinedMechanism}. Long transients dynamics were for
example also  suggested in a neural mass model of epileptogenic tissue
\cite{GOO12}.  However, another possible landscape  is that of a bi-stable
regime, for example, the dynamics of epileptic phenomena were determined from
statistics of such ictal transitions without occurrence of a prodrome phase and
DNBs \cite{SUF06}. 

For a DNB being possibly an {\em a posteriori} explanation for the prodrome
sets the aim expressed by the second difference bewteen DNB and classical
biomarkers (see Sec.~\ref{sec:dnb}) into a broader perspective.  While in
complex diseases with sudden deterioration phases the aim of DNB is to detect
the pre-disease state, now in chronic disorders with episodic manifestations
the question is what is the neural correlate, i.e., what is the DNB, of the
prodrome? The aim could still be to more reliably detect the prodrome stage.
However, migraine patients, for example, already very often can correctly
predict upcoming attacks by their premonitory prodromal symptoms\cite{GIF03}.
Thus, the broadened aim is to use the understanding of the slow and still
gradual physiological changes in the prodrome to prevent the imminent abrupt
attack, that is, to quantify the prodrome by DNB to construct  a biofeedback
signal for therapeutic control techniques.

\begin{figure*}[t] \begin{center}
\fbox{\includegraphics[width=\textwidth]{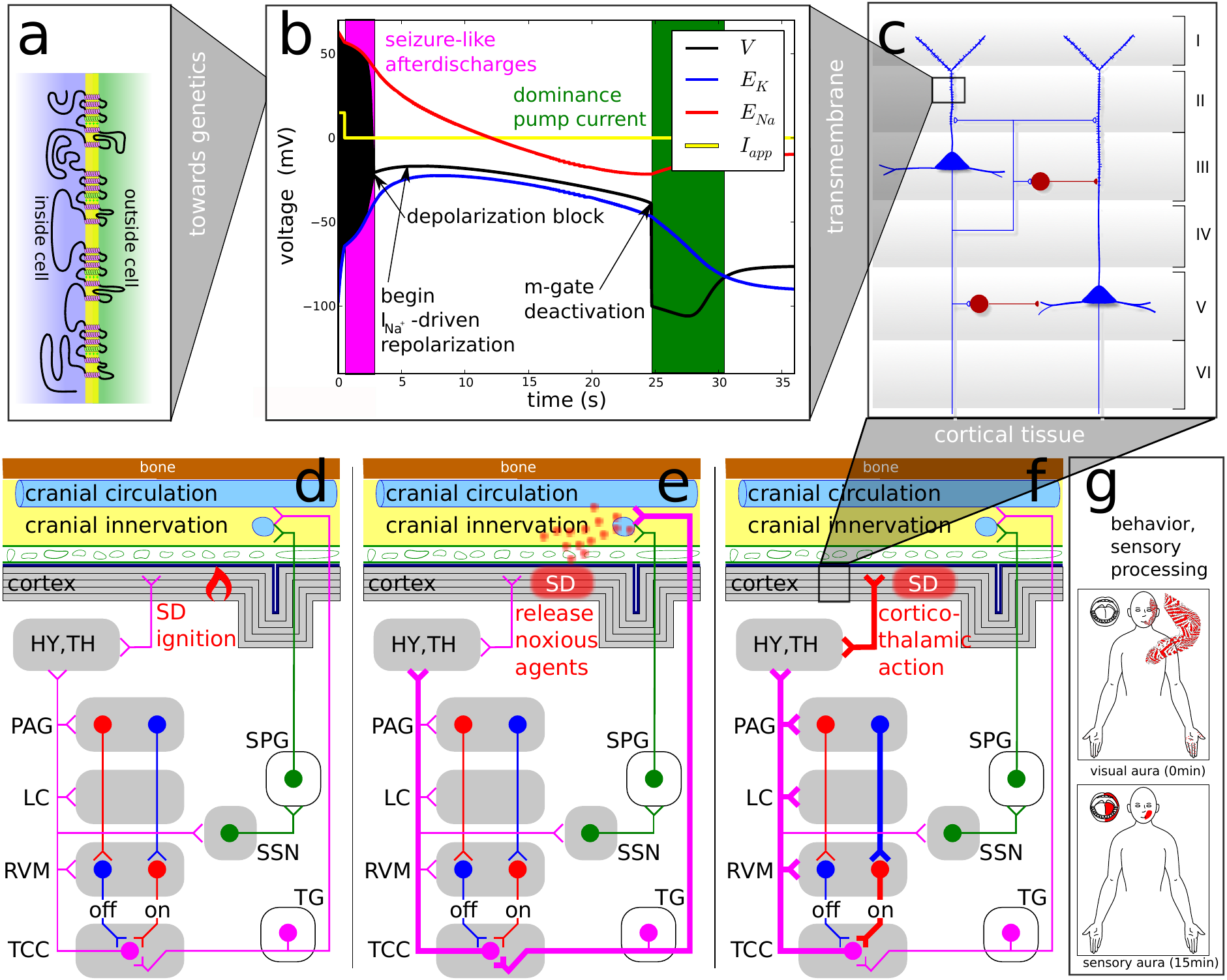}}
\end{center} \caption{\label{fig:fromCell2MGN}Migraine affects neural
activities across all levels.  (a) Rare monogenic subtypes exist and
genome-wide association study have revealed susceptibility loci for migraine
with and without aura. (b) Disturbed functions of ion channel, proteins that
regulate them, and pumps can be modelled by ion-conductance-based models. (c)
Neuronal subpopulations and local circuits in specific brain regions, in
particular, the layered cortex have also been investigated. (d) Higher-order
networks of central nuclei and peripheral ganglia form the migraine generator
network and its dynamical state may change during the course of an attack
(e)-(f). (g) Sensory processing  can be perturbed in particular during the aura
phase and behavioral changes occur in the prodromal such as food craving that
can lead to false interpretation of triggers. (See main text for further
detail.) } \end{figure*}

\section{Primary headaches as examples}
\label{sec:phae}

The theoretical concept of early-warning signals is independent of its
application\cite{SCH09g} and the concept of DNB is actually a model-free
concept \cite{CHE12a,LIU12b}. Still specific computational models are needed to
approach the broadened aim of therapeutic strategies. We cannot outline the
potential use of DNB as early-warning signals in general terms and therefore
focus on primary headaches and in particular on migraine.  Migraine patients
can very often correctly predict with premonitory symptoms their
attacks\cite{GIF03} and various neuromodulation methods are today available to
treat headaches\cite{MAG12}. Furthermore, computational models of migraine
exist and describe the complex processes on a wide range of time and space
scales, see Fig.~\ref{fig:fromCell2MGN}.  

Migraine is a primary headache disorder characterized by recurrent episodes of
head pain, often throbbing and unilateral sometimes preceded by neurological
symptoms called aura. The known pathophysiologic mechanism suggests that this
disorder should be considered as a neurovascular headache \cite{CHA12}.
Migraine is a highly disabling disorder being according to the World Health
Organization the seventh most disabling in the world and the fourth most
disabling among women. Cluster headaches are also a primary headache disorder
and the brainstem structures and subnetworks discussed below are also involved
in cluster headache so that these concepts can likely be applied in some form
to cluster headache, too.

Note that migraine is a chronic disorder with episodic manifestation, also
referred to as episodic migraine, while the relative new term 'chronic
migraine' (CM) refers  to chronification of the recurrent attacks, that is,
more then 15 headache days per month over a 3 month period, at least 8
migrainous, and in the absence of medication overuse.

In migraine without aura (MO), attacks are usually associated with nausea,
vomiting,  sensitivity to light (photophobia) and sound (phonophobia), and
attacks are worsened by movement. Migraine with aura (MA) involve in
addition—but also rarely exclusively—neurologic symptoms (called aura)
\cite{OLE13}. Migraine aura symptoms are most often visual field disturbances,
but affect also other sensory modalities or cognitive functions \cite{VIN07}.
Aura symptoms are caused by a wave of cortical spreading depression. Whether
this wave also contributes to the headache phase in migraine is currently
debated \cite{KAR13}.  Epidemiological studies suggest that MA and MO share the
same headache phase, however, on average the pain seems to be less severe and
shorter lasting in MA than in MO \cite{RAS92}.  

The migraine cycle can be divided in four distinct stages and the headache-free
interictal interval. The prodrome is characterised by feeling tired and weary,
difficulty concentrating, yawning, polyuria, sweating, and fatigue are some
typical features \cite{GIF03}.  It lasts up to one day and occurs in about 60\%
of the reported cases.  During the aura phase,  all kind of sensory
disturbances, visual most frequently, are reported. This phase lasts 5-60min
and occurs in about 30\% of the reported cases.  Migraine headache is often
unilateral and throbbing---though interestingly from a dynamical systems
theoretical point of view, this seems to be not related to blood pulsation but
is a signature of rhythms in the brain \cite{AHN10,MO13}.  Migraine headaches
last 4-72 hours and occur in about 95\% of the cases.  Last,  the postdrome is
characterized by tiredness, difficulty concentrating, weakness, dizziness,
persistence of sensitivity to light and noise, lethargy, and fatigue, together
often described as `headache hangover'. It lasts a few days and occurs in about
70\% of the reported cases.  Together with the attack-free interval these
phases compose the migraine cycle with associated severe disabilities and
reduced quality of life even between attacks \cite{BRA08}.

\section{Migraine theories} \label{sec:mt}

While we expect that quantitative methods such as computational neuroscience
will become increasingly important in migraine research to keep up with the
proliferation of imaging data and to confine current theories,  it was the
advent of non-invasive imaging that changed the view on this headache type in
the first place with some  intriguing qualitative observations. In the 1980s
and into the mid-1990s, it became clear by studies of the clinical features of
migraine and its physiological mechanisms that there is no clear correlation
between blood flow changes and the headache but instead imaging studies
suggested \cite{WEI95a} that upstream events in the brainstem named `migraine
generator' (MG) \cite{WEL01} lead to the cascade resulting in migraine pain.
Furthermore, imaging migraine with aura revealed that the phenomenon of
cortical spreading depression (SD) is the neural correlate of migraine aura
\cite{OLE81,HAD01}.  To date, the MG and SD theories are facing each other but
are probably not irreconcilable.

In brief, the MG theory states that a dysfunction of the brainstem causes the
pain and via further connections (directly and indirectly) the brainstem is
also responsible for the neurological changes before the headache phase
including symptoms in the prodromal phase and the aura phase, in particular,
the MG is causing SD via a failure of vasomotor control
(Fig.~\ref{fig:fromCell2MGN}d).  The SD theory of migraine assumes that this
cortical wave causes the neurological systems (aura) \cite{LAU94} and it can
also cause the pain in the headache phase, which usually occur after the aura
phase, but cf.~Ref.~\cite{HAN12a}. In this theory, two independent pathways of
increased pain traffic are suggested, a peripheral and a central pathway. The
headache pain can either be triggered by a local release of noxious or
inflammatory substances during the hyperemic phase of SD \cite{NOS10,KAR13}.
These substances are thought to be transmitted outward in the direction
perpendicular to the cortical layers into the pain sensitive meninges.  This
transmission results in an activation of pain receptors (nociceptors) and local
inflammation, which eventually triggers the migraine headache (peripheral
pathway, Fig.~\ref{fig:fromCell2MGN}e). Or the alternative explanation is that
SD leads to pain by a corticothalamic feedback that affects sensory processing
by third-order neurons arising within the sensory thalamus. This is supported
by the efficacy of calcitonin gene-related peptide receptor antagonists at the
level of these neurons in migraine therapy\cite{SUM10} (central pathway,
Fig.~\ref{fig:fromCell2MGN}f).  Probably, a combination of the peripheral and
central pathway leads to this complex disorder and quantitative models as
outlined below can help to disentangle the relative contributions and
dependencies.

\section{Computational models in migraine} \label{sec:cm}

Computational models of spreading depression date back to the mid 1970s. A
simple rule-based cellular automata model was used to mimic the two-dimensional
spread on the cortical surface, including spiral-shaped SD waves reentering
around a functional block \cite{RES75} that can also be observed experimentally
\cite{DAH97}.  Partial differential equations were used to describe the
detailed physiological cellular events \cite{TUC78}. For a detailed review of
modelling SD including the mathematical description see the excellent review by
Miura et al. \cite{MIU07}. In the following, only basic approaches focusing on
the principles of cellular dynamics and tissue pattern formation in SD are
introduced. We start with cellular dynamics of SD and continue with the events
in cortical tissue. The mathematical equations are given in an Appendix.

\subsection{Cellular dynamics in SD} \label{sec:cell}

The focus in computational models of the cellular dynamics in SD is on
transmembrane events that can be observed in {\em in vitro} experiments
\cite{SOM01}.  The advantage of such experiments is that the phenomenon of SD
can be studied in isolation and the complex vasomotor feedback involved in SD
can be neglected\cite{AYA13}, though current modelling approaches include
metabolic and perfusion effects on SD\cite{CHA12a}. With this local focus set,
SD is essentially a massive but temporary perturbation of ion homeostasis.  The
ion concentrations are under physiological conditions  kept within a narrow
range  and their gradient across the cell membrane is treated in traditional
computational models of Hodgkin-Huxlex (HH) type as a constant battery, the
reversal potential or Nernst potential \cite{HOD52}.  During SD, however, ion
concentrations can change by over one order of magnitude, so that the original
HH approach to describe the potential changes across the excitable membrane
must be extended.  For instance, during SD the extracellular potassium
concentration raises from 3mM within a few seconds to peak values around 55mM,
plateaus at this level for about 30 seconds and recovers to the physiological
range within one to two minutes \cite{LAU94}.  Furthermore, neuronal activity
is depressed due to the nearly complete elimination of the transmenbrane ion
gradients, in other words, the cells release almost all of their available
energy---the Gibbs free energy---the supply for  physiological neural activity
\cite{DRE12}. This depressed activity state can spread throughout gray matter
in the brain, hence this activity is named as SD (see below the section on
spatial patterns forming in cortical tissue).

To model local cellular events in SD, Kirchhoff's current law is used as first
introduced in the HH framework\cite{HOD52}.  The major extension, however, is
that in addition to the original transmembrane potential $V$ and the variables
activating and inactivating conductances of transmembrane channels
(voltage-gating variables, usually called $m$, $n$, and $h$) further functions
and dynamic variables for the ion homeostasis are introduced.  First, an ATP
driven ion pump is modeled that exchanges intracellular sodium ions with
extracellular potassium at a $3/2$ ratio and therefore changes not only ion
concentrations dynamically but also leads to a depolarizing transmembrane
current $I_{pump}$. Second, extracellular and intracellular ion concentrations
become dynamic variables and thus increase the phase space of the model.  We
refer to the Appendix A for the corresponding mathematical equations. Such an
extended framework is able to model the long lasting dynamics on time scales of
minutes, while, for example, a single action potential, which is described in
the first generation HH-type framework, lasts about a milli\-second, a time
scale separation of about five orders of magnitude.  This new time scale
separation is the fundamental characteristic of this second generation HH-type
framework. 

The first generation of HH-type models for single cells are also called
conductance-based models. In the parallel development of neural mass models for
cell populations one often refers to rate- or activity-based modelling. In each
case, the model is named after the respective dependent dynamical variables
other than the transmembrane potential. In this terminology, the second
generation HH-type models necessary to model SD are ion-conductance-based
models. Note that ion-conductance-based models are necessarily
`tissue-embedded' single cell models because the amount of cell membrane
surface area per unit tissue volume, in short, the surface area-to-volume ratio
plays a significant role. While this factor does not affect HH type
conductance-based models, in ion-conductance-based models it gives rise to the
long time scale of SD excitation cycles as compared to much shorter actions
potentials and may likewise cause slow modulations of activity in extended
rate- or activity-based neural mass models.

With an ion-conductance-based model of a total of 9 dynamical variables the
main features of SD that happen across the cell membrane can be modelled (see
Fig.~\ref{fig:fromCell2MGN}b). Using electroneutrality and further symmetry
constraints,  ion-conductance-based models can be further reduced and become
amenable  for bifurcation  and geometrical phase space analysis. On the other
end towards physiologically more realistic ion-conductance-based models of SD,
several dozens of dynamical variables in various electrically coupled neural
compartments ($\sim$200, which effectively multiplies the number of dependent
dynamical variables)  can be modelled in  tissue-embedded single cell
models\cite{KAG00,SHA01,MIU07,SOM08}. 

Although in reduced ion-conductance-based models the time course of the dynamic
variables  can be to some extent distorted as compared to more realistic models
\cite{KAG00,SHA01,MIU07,SOM08}, main features of SD are revealed
(Fig.~\ref{fig:fromCell2MGN}b).  Two phases and three points stand out because
of their significant mechanistic meaning in SD as an excitation phenomenon: the
seizure-like afterdischarges, the subsequent excitation block, the begin of a
$I_{Na^+}$-driven transient repolarization, the $m$-gate deactivation that
finally initiates a phase where the pump current $I_{pump}$ can become
dominant, see Fig.~\ref{fig:fromCell2MGN}b. 

Of all these events highlighted in Fig.~\ref{fig:fromCell2MGN}b most notably is
the recovery phase of SD at around  $\sim$24.5s.  According to the
computational models, recovery is due to a deactivation of the sodium activator
$m$ driven by a dominate $I_{Na^+}$ that actually repolarizes $V$ shortly after
the seizure-like firing stops due to a depolarization block.  After the
deactivation of the sodium activator $m$, the pump current $I_{pump}$   drives
the membrane potential $V$ (black) for several seconds outside the window of
the two reversal potential of sodium and potassium, $E_{Na^+}$ (red) and
$E_{K^+}$ (blue), respectively. The ignition of SD is due to a torus
bifurcation (not shown, unpublished results) of the tonic seizure-like firing
caused by the initial stimulation phase. This suggests that after the offset of
the stimulation at 0.5s the state of the dynamical system is already behind a
canard-like trajectory that serves as a threshold causing firstly a brief phase
of sustained (2-3s) afterdischarges and then a long lasting recovery phase
during which the system is refractory.

Next, we consider large-scale tissue events caused by SD. But before that it is
worthwhile noting that cellular models often do not include lateral space,
except for the original work by Miura and Tuckwell \cite{TUC78} and some recent
approaches from the group of Miura \cite{YAO11a} and an isolated work by
Shapiro \cite{SHA01} considering osmotic forces and gap junctions. The
compartments mentioned before extend in the vertical direction to model the
apical dendritic tree of pyramidal cells in the cortex \cite{KAG00,SOM08}.  But
the very detailed models by Kager, Wadman, and Somjen are not spatially
extended to describe the lateral extension, that is, they describe SD not as an
excitable phenomenon in a spatially extended system, called excitable media. 

The lateral extension of cellular SD models is far from straight forward.
Naively adding diffusion to the extracellular dynamical ion concentrations
yields some insights but it does not reflect the necessary detail of neural
microcircuits that needs to be considered to take the spatial continuum limit
(Fig.~\ref{fig:fromCell2MGN}c). Neural field models, for example, describe this
limit for rate- or activity-based neural mass models\cite{BRE12}. The problem
is that the neural tissue is heterogeneous and the cell's embedding needs to be
considered in quite some detail for a lateral extension of cellular SD models.
A bottom-up approach toward an ion-conductance-based medium model that lacks
the support of an effective medium theory cannot expand from membrane to tissue
level in a canonical way and therefore loses its advantage over the simpler
top-down approaches.  Likewise, a coarse-graining in time from individual
spiking to rate or activity-based dynamics in simulated cell populations is
needed for the relevant time scales in migraine. A first promising approach was
made with an ion-conductance-based model in seizure spread\cite{ULL08}.

\subsection{Tissue events caused by SD} 
\label{sec:tissue}

The Gibbs free energy-depleted and activity depressed state of SD described in
the former section can spread through gray matter tissue by electrodiffusion
due to newly formed ion gradient along the membranes in the neuronal and glial
syncytium and the interstitial volume.  In accordance with noninvasively imaged
SD progression using high-field functional magnetic resonance imaging (fMRI)
\cite{HAD01} and reported visual field defects \cite{DAH08d}, a macroscopic
computational model of SD (see Appendix B)  was proposed in which to each
episodic migraine attack a particular spatio-temporal SD pattern is
formed\cite{DAH12b,DAH13}.  With this model a mechanism was suggested to
explain the shape of SD in migraine as a discontinuous wave segment that
spreads out in only one direction.  In a nutshell, the mechanism is based on
dissipative structures as self-organized localized solitary
patterns\cite{AKH08,KER94} that become transient objects with critical slowing
down---from a mathematical point of view a mechanism similar to the one of
DNB---that is modulated by cortical folding, see
Fig.~\ref{fig:combinedMechanism}.

These are the tissue events of SD that are of clinical importance for migraine
because we predict that they will determine not only the migraine aura but also
the headache phase for two reasons.  First, the spatio-temporal patterns of SD
determine via the peripheral pathway (Fig.~\ref{fig:fromCell2MGN}e) the
transmission of noxious or inflammatory mediators in the direction
perpendicular to the cortical surface into the pain sensitive meninges. This
transmission must be significantly convergent to reach noxious threshold
concentrations, whereby likewise the considerable heterogeneity of increased
nociceptor density around pial arteries and dural venous sinuses can interfere
with the distinctive self-organized patterns in the folded cortex. Second, the
particular spatio-temporal pattern of SD probably also determines via the
central pathway the specific corticothalamic feedback that affects sensory
processing in terms of sensitisation by third-order neurons
(Fig.~\ref{fig:fromCell2MGN}f).  We suggested that the two major migraine
subtypes, MO and MA, depend on the spatial SD patterns and the bifurcation that
is responsible for their transient nature one being without critical slowing
down and one showing  critical slowing down, respectively \cite{DAH12b,DAH13}.

The major advantage of generic, i.e., low dimensional, reaction-diffusion
models of SD lies in the fact that they provide insight in the phase space
structure of the whole class of models they represent.  In fact, localized
solitary patterns are not the only SD waves in the human cortex of clinical
relevance.  Re-entrant SD waves are believed to have the potential to worsen
stroke outcome in incremental steps with each wave circling near the infarct
core, while the same SD wave in the penumbra zone far from the infarct tissue
could also have a beneficial component by stimulating blood flow \cite{STR07}.
The infarct core is a static anatomical block, whereas spiral waves and
localized SD waves in migraine have a variable core (functional block) and
variable left-off tissue, respectively \cite{DAH97,DAH99}.  These and other
clinically important spatial pattern forming mechanisms can be understood in
terms of phase space structures. These structures describe the migraine
aura--ischemic stroke continuum as transitions of two-dimensional wave pattern
in SD from retracting localized (migraine) to stationary waves (persistent
migraine aura without infarct) to re-entrant waves (stroke) \cite{DAH09a}.
Therefore, the precise spatio-temporal SD pattern formation is of immense
clinical interest \cite{DRE11}. 

To simulate the clinically observed patterns in migraine\cite{HAD01,DAH08d}, a
generic reaction-diffusion mechanism can be described in abstract terms of
activator-inhibitor kinetics.  In fact, simple activator-only kinetics date
back to a mechanism of SD described in 1963 by Grafstein based on an otherwise
unpublished suggestion by Hodgkin and  Huxley, to date called the
Grafstein-Hodgkin-Huxley (GHH) model \cite{GRA63}.  The GHH model assumed that
the extracellular potassium ion concentration is an activator in SD. We
extended this scheme by introducing two inhibitory mechanisms, an immobilized
inhibitor and a mean field inhibition (for details of the full model see
Ref.~\cite{DAH12b}).

The abstract activator and inhibitor variables are lump variables. Unlike in
the original GHH model, it is not necessary or even possible to identify
particular physiological quantities, like $[K^+]_e$, with them. Instead, the
activator should be viewed as a positive feedback element with a bistable
energy state. The activator and inhibitor or rather the respective networks of
dynamical variables that constitute these  elements can be described in great
physiological detail on the cellular level (see the former section) yet they
are also new emerging conceptual quantity that allows us to describe the
phenomenon of SD directly on the macroscopic level.  The high energy state (a
stable fixed point) corresponds to the maintenance of homeostasis far from
thermodynamic equilibrium and the low energy state (another stable fixed point)
to the state where the cellular Nernst reversal potentials are eliminated, that
is, a state at or near the thermodynamic Donnan equilibrium \cite{DRE12}.  In
the presence of inhibitors as recovery variables, which were not included in
the GHH model, the low energy stable fixed point becomes a transient state that
can, however, spread through gray matter tissue by electrodiffusion as a
depleted and activity depressed state.

\begin{figure}[t!]
\includegraphics[width=\columnwidth]{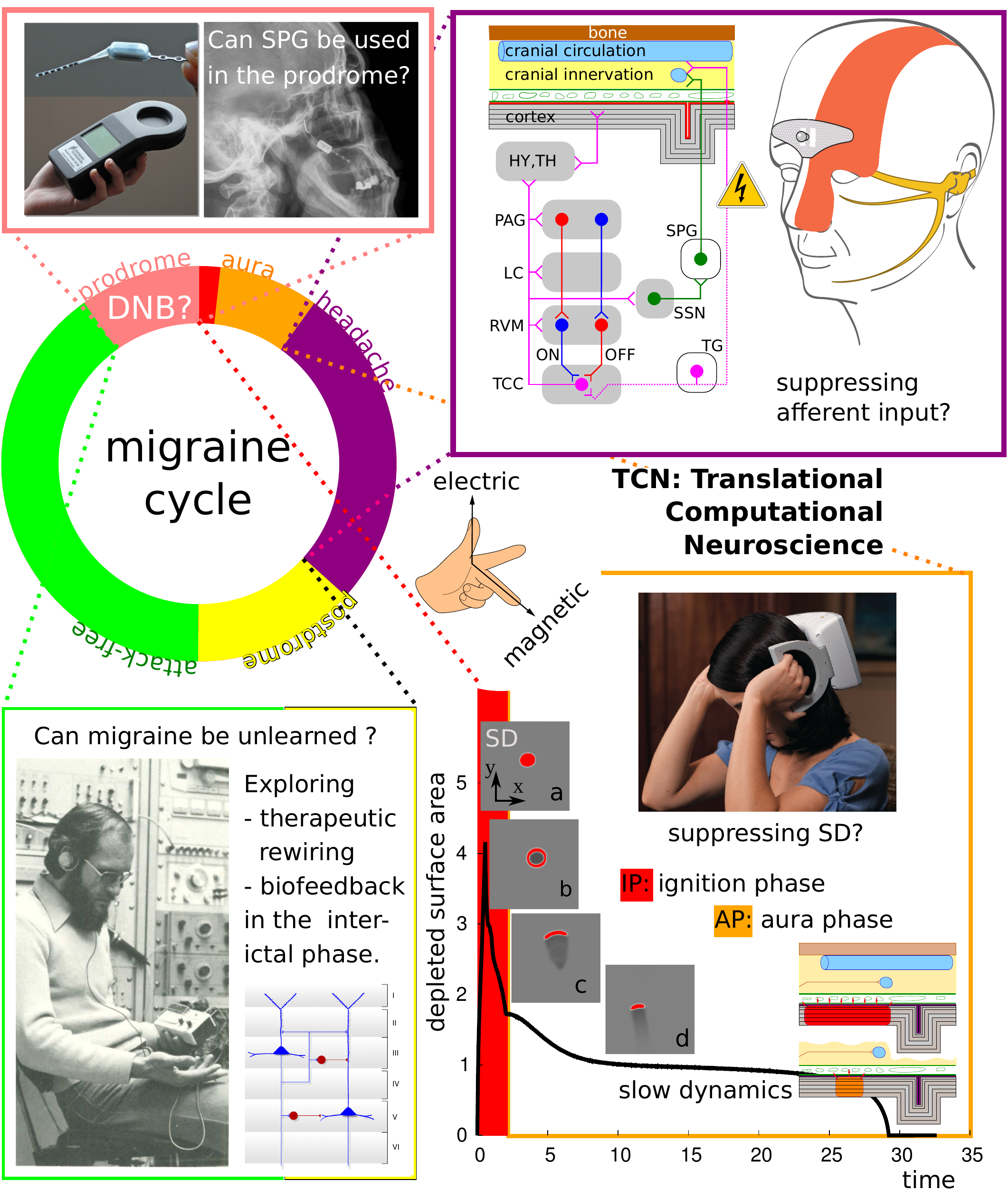}
\caption{\label{fig:neuromodulationDNB} Various neuromodulation techniques are
currently investigated in chronic migraine, but note that, as of to date, there
is no approved indication for neuromodulation with electric or magnetic
stimulation for episodic migraine. We suggest to investigate these techniques
also in episodic migraine and to use quantitative models and modern control
theory to determine an optimal stimulation protocol, target structure and phase
within the migraine cycle. Photograph top left (SPG stimulation): courtesy
Autonomic Technologies, Inc.\ \copyright{} 2013. Inset figure top right:
courtesy of STX-Med \copyright{} 2013. Photograph bottom right (TMS device):
courtesy of Neuralieve, Inc.\  \copyright{} 2013 (trading as eNeura
Therapeutics).  } \end{figure}

\section{Concluding remarks on translating theoretical concepts into neuromodulation applications}
\label{sec:nm}

The design of new neuromodulation strategies in migraine requires to understand
how electric and magnetic stimulation changes neural activity. Even more
important are quantitative models of the network dynamics  during this
neurological condition to optimize stimulation protocols. In
Sec.~\ref{sec:cell}, we described cellular dynamics that are key to understand
how electric and magnetic stimulation influences brain activity. The cellular
dynamics are also the microscopic foundation of self-organized macroscopic
tissue events, which are of direct clinical relevance. In
Sec.~\ref{sec:tissue}, we show that on the macroscopic level SD manifests as a
localized wave in the human cortex during migraine with aura. From the model we
can predict that the macroscopic spatio-temporal SD patterns (shape, size, and
duration) determine to some extent the clinical manifestation of a migraine
attack \cite{DAH13}.

As we argued in Sec.~\ref{sec:mt}, the spatio-temporal SD events can determine
the amount of noxious input into the pain sensitive structures
(Fig.~\ref{fig:fromCell2MGN}e) as well as modulate corticothalamic feedback
loops (Fig.~\ref{fig:fromCell2MGN}f).  However, the question  about the most
upstream event in the course of a migraine attack that would directly lead to
the optimal target of preventative intervention is probably too naive.  The
ignition of SD is currently unknown but a failure of vasomotor control from the
brainstem, which is part of the MG theory, is a likely scenario
(Fig.~\ref{fig:fromCell2MGN}d).  It further is plausible that these SD waves
will in turn also affect the brainstem activity because SD itself causes a
complex vasomotor reaction \cite{AYA13}.  This allows only one conclusion:
Migraine is an inherently dynamical disease with a complex network of
interdependent events rather than a disease with a linear course from upstream
to downstream events, a course that is also called into question by a rather
variable cycle of recurrent episodic symptoms in migraine with phases such as
the aura that occurs only in a fraction of the individual attacks \cite{GOA02}.  

The central question therefore is about different brain structures and in which
phase of the migraine cycle they are promising neuromodulation targets
(Fig.~\ref{fig:neuromodulationDNB}) in the framework of complex dynamical
systems and control theory\cite{AIH10}.  We still lack a unified quantitative
model combining the events in SD theory and MG theory  of migraine
(Sec.~\ref{sec:mt}) to interpret their relation, simulate the impact of
potential neuromodulation intervention, and make concrete predictions about
optimized stimulations protocols that can be tested.  In this situation,
however, the theoretical framework of DNB provides a sound basis to address
these challenges because DNB are based on a model-free concept and, if DNB can
be identified in migraine, they will help to develop a unified theory.  Namely,
DNB will not only show whether a symptom is approaching or not, but also help
to construct such a unified model explaining how the symptom appears, and
providing a feedback-control-scheme to suppress the symptom.

We end this review by introducing briefly two neuromodulation techniques that
are currently investigated in migraine in two different phases of the migraine
cycle targeting two different brain structures. Then, we speculatively suggest
a promising third neuromodulation technique that may be of particular use
during the prodromal phase that could reflect a DNB (cf.
Sec.~\ref{sec:dnbcdem}).  Furthermore, we also suggest taking the attack-free
interval into account for prophylactic intervention.

Note that, as of to date, there is no approved indication for neuromodulation
with electric or magnetic stimulation for episodic  but only for chronic
migraine ($>$15 headache days per month). In fact, due to the severeness of
pain, most neuromodulation studies in headache research are or have been
conducted among patients with refractory cluster headache. However, the field
is rapidly increasing \cite{MAG12} and given that newer devices require
noninvasive or minimally invasive implant surgery involving a new generation of
intelligent control methods  also less disabled patients with high frequency
epsisodic migraine with unsatisfactory treatment response could benefit from
efficient neuromodulation.

\subsection{Transcutaneous electrical nerve stimulation}

Transcutaneous and transcranial (see the next subsection) stimulations are the
two noninvasive modes currently at hand.  The transcutaneous electrical nerve
stimulation (TENS) modulates neural activity by targeting noninvasively
peripheral nerves.  In migraine, a supraorbital transcutaneous stimulation
(STS) was tested with a portable device that looks like a silver headband and
targets the ophthalmic division of the trigeminal nerve  that terminates in the
central trigeminal nucleus caudalis (TNC) and the C1-C2 regions of the cervical
spinal cord, together known as the trigeminocervical complex (TCC), see top
right in Fig.~\ref{fig:neuromodulationDNB}. 

In a recent study, the stimulation was tested with biphasic rectangular AC
impulses at 60Hz.  The therapeutic gain (26\%) is within the range of those
reported for other preventive drug and nondrug antimigraine
treatments\cite{SCH13d}.  The mode of action or even quantitative methods to
test paradigms are currently missing, though it seems plausible that if central
sensitization of second-order neurons is causing migraine pain, this method is
suppressing  the inbound sensory traffic which could add to sensitization, and
it should therefore be used in the acute headache phase rather than
prophylactically.

\subsection{Transcranial magnetic and electrical stimulation}

The transcranial stimulation targets not peripheral neurons but the cortex. Two
noninvasive neuromodulation techniques are available in migraine, transcranial
magnetic stimulation (TMS) and  transcranial electrical stimulation (TES).  A
recent carefully sham-controlled larger trial with a portable TMS device was
undertaken as a promising noninvasive neurostimulator for disrupting SD early
in the aura phase \cite{LIP10}, see bottom right in
Fig.~\ref{fig:neuromodulationDNB}. The results (effective in aborting migraine
attacks) of this sham-controlled study were promising, i.e. it seems  to
disrupt SD, but that would limit its use to the acute aura phase. However,
considering the concept of silent aura \cite{AYA10} and the mathematical model
for SD \cite{DAH12b},  TMS could potentially be effective when used in the very
early headache phase targeting the shorter lasting phase of SD. In either case,
there is a clear need for further studies and  the optimization of the
stimulation protocol for targeting localized spatio-temporal reaction-diffusion
patterns \cite{DAH12b}.

The simplest technology with the least safety issue is probably TES.  Two
versions of TES exist, the transcranial direct current stimulation (tDCS), in
which cathodal tDCS inhibits while anodal tDCS increases neuronal firing. The
alternating current stimulation (tACS) has not yet been clinically tested in
migraine.  Preliminary evidence was found for patients with chronic migraine
having a positive, but delayed, response to tDCS applied to motor (anodal) and
oribitofrontal (cathode) cortices\cite{DAS12}. In addition, computational
simulations of current flow through brain regions were used to interpret the
effects in this study. With the help of such simulations, a principal
understanding of pain modulation can be gained, whether this modulation is due
to the effect of TES on SD or on the dynamics of the pain matrix and deeper
structures including thalamus and brainstem.  Effective prophylactic therapy in
migraine was also studied with cathodal tDCS over the visual cortex with the
anode overlying the motor and sensory cortices \cite{ANT11a}.  

\subsection{Sphenopalatine ganglion stimulation}

Beside the two mechanism explained by the SD theory
(Fig.~\ref{fig:fromCell2MGN}e,f), also other mechanisms could activate pial and
dural nociceptors and trigger the pain of the headache phase of migraine. It
was suggested by Burstein and Jakubowski \cite{BUR05a} that a common pathway
for various  migraine triggers could be provided by the pre- and postganglionic
parasympathetic neurons in the superior salivatory nucleus (SSN) and
sphenopalatine ganglion (SPG). A number of projections converging in the SSN
seem to be  coming from brain areas that can explain typical migraine triggers,
for example  stress, dietary, and external stimuli such as smell. Targeting the
SSN through antidrome or feedback mechanisms of the SPG could therefore be a
potential target in migraine (prevention).

Interestingly, it was also argued that for many if not all of the typical
triggers the role of cause and effect is actually reversed and premonitory
symptoms are  mistaken as triggers \cite{GOA13a,HOU13}.  For example, eating
chocolate, as a typical food craving symptom before the headache phase of
migraine, is a brain-driven response within the premonitory phase and not a
trigger, i.e. the cascade of the attack has already started and the only
symptom at that stage is the craving with all the other symptoms still to come.
Likewise, sensitivity to light (photophobia) during the premonitory phase could
mistakenly identified as a trigger by light.   In the context of DNB this is
certainly plausible as this concept predicts a common subnetwork such as SSN
and SPG to be involved in the premonitory symptoms.

Given the theory that the pre- and postganglionic parasympathetic neurons in
the superior salivatory nucleus (SSN) and sphenopalatine ganglion (SPG) could
be a common pathway for migraine triggers \cite{BUR05a}, minimally invasive
sphenopalatine ganglion stimulation might be suited in particular in the
prodromal phase in episodic migraine.  An very innovative  implantable
battery-free SPG stimulator has been developed to apply on-demand stimulation
for the treatment of migraine and other primary headache\cite{PAE12}.  In a
pilot study (n=11), the potential of electrical stimulation of the SPG has
already  been shown in attack abortion, 2 patients were pain-free within 3
minutes of stimulation, 3 had pain reduction, 5 had no pain relief, 1 was not
stimulated\cite{TEP09}. However, in this pilot study, the headache was allowed
to intensify up to 6 hours before stimulation was initiated, where SPG
stimulation might theoretically be less efficient to influence a migraine
attack than it could have in the prodromal phase. More to the point, a recent
paper described not only a marked abortive but also a preventative effect in
cluster headache patients \cite{SCH13h} which could best be explained by
feed-back mechanisms from the SPG to the SSN and thus underlining the above
mentioned theory and pointing towards the SPG as a suitable target for attack
prevention. However, all this is based on a theoretical concept and it should
be noted that in migraine research no valid studies have been published yet to
prove that SPG stimulation could have a positive effect in this syndrome as
well. \\

To summarize, already to date various neuromodulation methods for treatment of
primary headache have been studied, beside the above mentioned also other
invasive techniques such as deep brain stimulation and minimally invasive
occipital nerve stimulation \cite{MAG12}.  To fully realize the potential of
computational models that can be translated into a new neuromodulation paradigm
a modern control theory is needed.  The challenge in the future will be to
integrate the known mechanisms into a tractable theoretical migraine model, but
also  to obtain human data from clinical studies and identify DNB. Therefore,
new clinical study designs are needed to investigate the macroscopic dynamics
in migraine. For example, fMRI brainstem specific ROI-based imaging paradigms
and fluorodeoxyglucose positron emission tomography (FDG-PET) combined with
magnetic resonance imaging (MRI) could identify important neuronal centers in
the brainstem and describe their connectivity pattern \cite{SON12}. Such
imaging methods together with quantitative methods and theoretical concepts are
important to characterize dynamics in the migraine generator network and
identify DNB over several migraine cycles in controls and patients suffering
from episodic migraine.  \\

\section{Appendix:  Model equations}

\subsection{Cellular models}

To model local cellular events during SD, we start with Kirchhoff's current law
that was introduced as an equivalent membrane circuit by Hodgkin and Huxley
\cite{HOD52}
\begin{eqnarray} \label{eq:kirchhoff}
\frac{\mathrm{d}V}{\mathrm{d}t} = -\frac{1}{C_m}(I_{Na^+}+I_{K^+}+I_{Cl}+I_{pump}-I_{app}).
\end{eqnarray}
The variable $V$ is the potential difference or voltage across the membrane
relative to the grounded extracellular side. The cell membrane has a capacity
$C_m$. The currents $I_{\text{ion}}$ are the transmembrane currents with {\em
ion} $\in \{Na^+,K^+,Cl^-\}$. $I_{app}$ is an externally applied current.

In the original HH framework, an unspecified leak current was modelled but
not a chloride current. The major extension, however, is the pump current
$I_{pump}$, which can be modeled for example as\cite{CRE08}:\begin{widetext}
\begin{eqnarray}
I_{pump}(\mathit{Na}_i,K_e)&=&\rho\bigg(1+\exp{\bigg(\frac{25-\mathit{Na}_i}{3}\bigg)}\bigg)^{-1}\bigg(1+\exp{(5.5-K_e)}\bigg)^{-1}.
\end{eqnarray}\end{widetext}
The pump current $I_{pump}$ with pump rate $\rho$ describes the ATP driven
exchange of intracellular sodium ions with extracellular potassium at a $3/2$
ratio.

The  ion concentrations $[\text{ion}]$ change due to fluxes proportional to the
ion currents.  The additional rate equations for intracellular ion
concentrations are
\begin{eqnarray}
\frac{\mathrm{d}\,[\text{ion}]}{\mathrm{d}t} &=& -\frac{\gamma}{\omega}(I_{\text{ion}}+ \alpha I_{pump})\ ,
\end{eqnarray}
with $\alpha$ being $3$, $2$, and $0$ for $Na^+$,$K^+$, and $Cl^-$,
respectively.  The parameter $\gamma$ converts currents to ion fluxes and
depends on the membrane surface area $A_m$ and Faraday's constant $F$, $\gamma
= \frac{A_m}{F}$.  The parameter $\omega$ is the volume of the intracellular
space (ICS).   

In addition, also dynamical gating variables $n$, $h$ and $m$ have to be
introduced, as in the original HH model. The potassium activator, sodium
activator and sodium inactivator are defined by the voltage-dependent gating
dynamical variables  $n$, $m$ and $h$, respectively. These are given by 
\begin{eqnarray}
\frac{\mathrm{d}\,p}{\mathrm{d}t} &=& \frac{1}{\tau_p} \left( p_\infty - p\right), \label{eq:gating}
\end{eqnarray}
for $p \in \{n, m, h\}$. 

\subsection{Reaction-diffusion models}

To simulate the clinically observed patterns in migraine\cite{HAD01,DAH08d}, a
generic reaction-diffusion mechanism can be described in abstract terms of
activator-inhibitor kinetics of just two variables $u$ and $v$, respectively.
We extended the GHH model (see the main text) in the following way (also for
details of the full model see Ref.~\cite{DAH12b}):
\begin{eqnarray} 
\varepsilon \frac{\partial u}{\partial t} &=&  u -\frac{1}{3} u^3 -v + \nabla^2u,   \label{eq:u}\\ 
\frac{\partial v}{\partial t} &=& u + \beta + K \int\!\!\!\int H\left(u\right)\,\textrm{d} x\textrm{d} y, \label{eq:v}
\end{eqnarray} 
with $H$ being the Heaviside step function and the space coordinates are given
by $x$ and $y$. Both activator $u$ and inhibitor $v$ are lump variables (see
the main text).  We set the parameter as follows: time scale separation
$\varepsilon=0.04$, the threshold $\beta=1.32$, and the mean field coupling
$K=0.003$.

There are two stable solutions of this model, namely the homogeneous steady
state and the traveling wave. Moreover, there exists a saddle solution forming
a localized solitary pattern whose stable manifold is the basin boundary of the
stable solutions. By varying a control parameter, the stable traveling wave
solution and the saddle solution collide causing the saddle-node bifurcation.
The saddle-node bifurcation and the critical slowing down with this model
suggest the applicability of DNB to detect SD.

\section*{Acknowledgments} 

The authors kindly acknowledge the support of M.A.D. from  the
Bundesministerium f\"ur Bildung und Forschung (BMBF 01GQ1109) and of A.M. by
the German Research Foundation (SFB936/A5).  N.F., Y.H., and K.A. are supported
by the Aihara Innovative Mathematical Modelling Project, the Japan Society for
the Promotion of Science (JSPS) through its ''Funding Program for World-Leading
Innovative R\&D on Science and Technology (FIRST Program),`` initiated by the
Council for Science and Technology Policy (CSTP). The authors also kindly
acknowledge helpful discussions with Luonan Chen and Rui Liu.

\section*{Disclosure: Conflict of interest} M.A.D. received  honoraria  for
consulting services at Neuralieve Inc. (trading as eNeura Therapeutics).

\section*{References}

\begin{thebibliography}{73}%
\makeatletter
\providecommand \@ifxundefined [1]{%
 \@ifx{#1\undefined}
}%
\providecommand \@ifnum [1]{%
 \ifnum #1\expandafter \@firstoftwo
 \else \expandafter \@secondoftwo
 \fi
}%
\providecommand \@ifx [1]{%
 \ifx #1\expandafter \@firstoftwo
 \else \expandafter \@secondoftwo
 \fi
}%
\providecommand \natexlab [1]{#1}%
\providecommand \enquote  [1]{``#1''}%
\providecommand \bibnamefont  [1]{#1}%
\providecommand \bibfnamefont [1]{#1}%
\providecommand \citenamefont [1]{#1}%
\providecommand \href@noop [0]{\@secondoftwo}%
\providecommand \href [0]{\begingroup \@sanitize@url \@href}%
\providecommand \@href[1]{\@@startlink{#1}\@@href}%
\providecommand \@@href[1]{\endgroup#1\@@endlink}%
\providecommand \@sanitize@url [0]{\catcode `\\12\catcode `\$12\catcode
  `\&12\catcode `\#12\catcode `\^12\catcode `\_12\catcode `\%12\relax}%
\providecommand \@@startlink[1]{}%
\providecommand \@@endlink[0]{}%
\providecommand \url  [0]{\begingroup\@sanitize@url \@url }%
\providecommand \@url [1]{\endgroup\@href {#1}{\urlprefix }}%
\providecommand \urlprefix  [0]{URL }%
\providecommand \Eprint [0]{\href }%
\providecommand \doibase [0]{http://dx.doi.org/}%
\providecommand \selectlanguage [0]{\@gobble}%
\providecommand \bibinfo  [0]{\@secondoftwo}%
\providecommand \bibfield  [0]{\@secondoftwo}%
\providecommand \translation [1]{[#1]}%
\providecommand \BibitemOpen [0]{}%
\providecommand \bibitemStop [0]{}%
\providecommand \bibitemNoStop [0]{.\EOS\space}%
\providecommand \EOS [0]{\spacefactor3000\relax}%
\providecommand \BibitemShut  [1]{\csname bibitem#1\endcsname}%
\let\auto@bib@innerbib\@empty
\bibitem [{\citenamefont {Mackey}\ and\ \citenamefont {Milton}(1987)}]{MAC87}%
  \BibitemOpen
  \bibfield  {author} {\bibinfo {author} {\bibfnamefont {M.~C.}\ \bibnamefont
  {Mackey}}\ and\ \bibinfo {author} {\bibfnamefont {J.~G.}\ \bibnamefont
  {Milton}},\ }\bibfield  {title} {\enquote {\bibinfo {title} {{{D}ynamical
  diseases}},}\ }\href@noop {} {\bibfield  {journal} {\bibinfo  {journal} {Ann.
  N. Y. Acad. Sci.}\ }\textbf {\bibinfo {volume} {504}},\ \bibinfo {pages}
  {16--32} (\bibinfo {year} {1987})}\BibitemShut {NoStop}%
\bibitem [{\citenamefont {Milton}\ and\ \citenamefont {Jung}(2003)}]{MIL03a}%
  \BibitemOpen
  \bibfield  {author} {\bibinfo {author} {\bibfnamefont {J.}~\bibnamefont
  {Milton}}\ and\ \bibinfo {author} {\bibfnamefont {P.}~\bibnamefont {Jung}},\
  }\href@noop {} {\emph {\bibinfo {title} {Epilepsy as a Dynamic Disease}}},\
  Biological and Medical Physics Series\ (\bibinfo  {publisher} {Springer},\
  \bibinfo {year} {2003})\BibitemShut {NoStop}%
\bibitem [{\citenamefont {Scheffer}\ \emph {et~al.}(2009)\citenamefont
  {Scheffer}, \citenamefont {Bascompte}, \citenamefont {Brock}, \citenamefont
  {Brovkin}, \citenamefont {Carpenter}, \citenamefont {Dakos}, \citenamefont
  {Held}, \citenamefont {van Nes}, \citenamefont {Rietkerk},\ and\
  \citenamefont {Sugihara}}]{SCH09g}%
  \BibitemOpen
  \bibfield  {author} {\bibinfo {author} {\bibfnamefont {M.}~\bibnamefont
  {Scheffer}}, \bibinfo {author} {\bibfnamefont {J.}~\bibnamefont {Bascompte}},
  \bibinfo {author} {\bibfnamefont {W.~A.}\ \bibnamefont {Brock}}, \bibinfo
  {author} {\bibfnamefont {V.}~\bibnamefont {Brovkin}}, \bibinfo {author}
  {\bibfnamefont {S.~R.}\ \bibnamefont {Carpenter}}, \bibinfo {author}
  {\bibfnamefont {V.}~\bibnamefont {Dakos}}, \bibinfo {author} {\bibfnamefont
  {H.}~\bibnamefont {Held}}, \bibinfo {author} {\bibfnamefont {E.~H.}\
  \bibnamefont {van Nes}}, \bibinfo {author} {\bibfnamefont {M.}~\bibnamefont
  {Rietkerk}}, \ and\ \bibinfo {author} {\bibfnamefont {G.}~\bibnamefont
  {Sugihara}},\ }\bibfield  {title} {\enquote {\bibinfo {title}
  {{{E}arly-warning signals for critical transitions}},}\ }\href@noop {}
  {\bibfield  {journal} {\bibinfo  {journal} {Nature}\ }\textbf {\bibinfo
  {volume} {461}},\ \bibinfo {pages} {53--59} (\bibinfo {year}
  {2009})}\BibitemShut {NoStop}%
\bibitem [{\citenamefont {Chen}\ \emph {et~al.}(2012)\citenamefont {Chen},
  \citenamefont {Liu}, \citenamefont {Liu}, \citenamefont {Li},\ and\
  \citenamefont {Aihara}}]{CHE12a}%
  \BibitemOpen
  \bibfield  {author} {\bibinfo {author} {\bibfnamefont {L.}~\bibnamefont
  {Chen}}, \bibinfo {author} {\bibfnamefont {R.}~\bibnamefont {Liu}}, \bibinfo
  {author} {\bibfnamefont {Z.~P.}\ \bibnamefont {Liu}}, \bibinfo {author}
  {\bibfnamefont {M.}~\bibnamefont {Li}}, \ and\ \bibinfo {author}
  {\bibfnamefont {K.}~\bibnamefont {Aihara}},\ }\bibfield  {title} {\enquote
  {\bibinfo {title} {{{D}etecting early-warning signals for sudden
  deterioration of complex diseases by dynamical network biomarkers}},}\
  }\href@noop {} {\bibfield  {journal} {\bibinfo  {journal} {Sci. Rep.}\
  }\textbf {\bibinfo {volume} {2}},\ \bibinfo {pages} {342} (\bibinfo {year}
  {2012})}\BibitemShut {NoStop}%
\bibitem [{\citenamefont {Liu}\ \emph {et~al.}(2012)\citenamefont {Liu},
  \citenamefont {Li}, \citenamefont {Liu}, \citenamefont {Wu}, \citenamefont
  {Chen},\ and\ \citenamefont {Aihara}}]{LIU12b}%
  \BibitemOpen
  \bibfield  {author} {\bibinfo {author} {\bibfnamefont {R.}~\bibnamefont
  {Liu}}, \bibinfo {author} {\bibfnamefont {M.}~\bibnamefont {Li}}, \bibinfo
  {author} {\bibfnamefont {Z.-P.}\ \bibnamefont {Liu}}, \bibinfo {author}
  {\bibfnamefont {J.}~\bibnamefont {Wu}}, \bibinfo {author} {\bibfnamefont
  {L.}~\bibnamefont {Chen}}, \ and\ \bibinfo {author} {\bibfnamefont
  {K.}~\bibnamefont {Aihara}},\ }\bibfield  {title} {\enquote {\bibinfo {title}
  {Identifying critical transitions and their leading biomolecular networks in
  complex diseases},}\ }\href@noop {} {\bibfield  {journal} {\bibinfo
  {journal} {Sci. Rep.}\ }\textbf {\bibinfo {volume} {2}},\ \bibinfo {pages}
  {813} (\bibinfo {year} {2012})}\BibitemShut {NoStop}%
\bibitem [{\citenamefont {Liu}\ \emph {et~al.}(2013)\citenamefont {Liu},
  \citenamefont {Wang}, \citenamefont {Aihara},\ and\ \citenamefont
  {Chen}}]{LIU13a}%
  \BibitemOpen
  \bibfield  {author} {\bibinfo {author} {\bibfnamefont {R.}~\bibnamefont
  {Liu}}, \bibinfo {author} {\bibfnamefont {X.}~\bibnamefont {Wang}}, \bibinfo
  {author} {\bibfnamefont {K.}~\bibnamefont {Aihara}}, \ and\ \bibinfo {author}
  {\bibfnamefont {L.}~\bibnamefont {Chen}},\ }\bibfield  {title} {\enquote
  {\bibinfo {title} {{{E}arly diagnosis of complex diseases by molecular
  biomarkers, network biomarkers, and dynamical betwork biomarkers}},}\
  }\href@noop {} {\bibfield  {journal} {\bibinfo  {journal} {Med. Res. Rev.}\ }
  (\bibinfo {year} {2013})}\BibitemShut {NoStop}%
\bibitem [{\citenamefont {Liu}, \citenamefont {Aihara},\ and\ \citenamefont
  {Chen}(2013)}]{LIU13b}%
  \BibitemOpen
  \bibfield  {author} {\bibinfo {author} {\bibfnamefont {R.}~\bibnamefont
  {Liu}}, \bibinfo {author} {\bibfnamefont {K.}~\bibnamefont {Aihara}}, \ and\
  \bibinfo {author} {\bibfnamefont {L.}~\bibnamefont {Chen}},\ }\bibfield
  {title} {\enquote {\bibinfo {title} {Dynamical network biomarkers for
  identifying critical transitions and their driving networks of biologic
  processes},}\ }\href@noop {} {\bibfield  {journal} {\bibinfo  {journal}
  {Quantitative Biology}\ }\textbf {\bibinfo {volume} {1}},\ \bibinfo {pages}
  {105--114} (\bibinfo {year} {2013})}\BibitemShut {NoStop}%
\bibitem [{\citenamefont {Aihara}\ and\ \citenamefont {Suzuki}(2010)}]{AIH10}%
  \BibitemOpen
  \bibfield  {author} {\bibinfo {author} {\bibfnamefont {K.}~\bibnamefont
  {Aihara}}\ and\ \bibinfo {author} {\bibfnamefont {H.}~\bibnamefont
  {Suzuki}},\ }\bibfield  {title} {\enquote {\bibinfo {title} {{{T}heory of
  hybrid dynamical systems and its applications to biological and medical
  systems}},}\ }\href@noop {} {\bibfield  {journal} {\bibinfo  {journal} {Phil.
  Trans.~R. Soc.~A}\ }\textbf {\bibinfo {volume} {368}},\ \bibinfo {pages}
  {4893--4914} (\bibinfo {year} {2010})}\BibitemShut {NoStop}%
\bibitem [{\citenamefont {Schiff}(2010)}]{SCH10h}%
  \BibitemOpen
  \bibfield  {author} {\bibinfo {author} {\bibfnamefont {S.~J.}\ \bibnamefont
  {Schiff}},\ }\bibfield  {title} {\enquote {\bibinfo {title} {{{T}owards
  model-based control of {P}arkinson's disease}},}\ }\href@noop {} {\bibfield
  {journal} {\bibinfo  {journal} {Phil. Trans. R. Soc. A}\ }\textbf {\bibinfo
  {volume} {368}},\ \bibinfo {pages} {2269--2308} (\bibinfo {year}
  {2010})}\BibitemShut {NoStop}%
\bibitem [{\citenamefont {Terman}\ \emph {et~al.}(2002)\citenamefont {Terman},
  \citenamefont {Rubin}, \citenamefont {Yew},\ and\ \citenamefont
  {Wilson}}]{TER02}%
  \BibitemOpen
  \bibfield  {author} {\bibinfo {author} {\bibfnamefont {D.}~\bibnamefont
  {Terman}}, \bibinfo {author} {\bibfnamefont {J.~E.}\ \bibnamefont {Rubin}},
  \bibinfo {author} {\bibfnamefont {A.~C.}\ \bibnamefont {Yew}}, \ and\
  \bibinfo {author} {\bibfnamefont {C.~J.}\ \bibnamefont {Wilson}},\ }\bibfield
   {title} {\enquote {\bibinfo {title} {{{A}ctivity patterns in a model for the
  subthalamopallidal network of the basal ganglia}},}\ }\href@noop {}
  {\bibfield  {journal} {\bibinfo  {journal} {J. Neurosci.}\ }\textbf {\bibinfo
  {volume} {22}},\ \bibinfo {pages} {2963--2976} (\bibinfo {year}
  {2002})}\BibitemShut {NoStop}%
\bibitem [{\citenamefont {Rubin}\ \emph {et~al.}(2012)\citenamefont {Rubin},
  \citenamefont {McIntyre}, \citenamefont {Turner},\ and\ \citenamefont
  {Wichmann}}]{RUB12}%
  \BibitemOpen
  \bibfield  {author} {\bibinfo {author} {\bibfnamefont {J.~E.}\ \bibnamefont
  {Rubin}}, \bibinfo {author} {\bibfnamefont {C.~C.}\ \bibnamefont {McIntyre}},
  \bibinfo {author} {\bibfnamefont {R.~S.}\ \bibnamefont {Turner}}, \ and\
  \bibinfo {author} {\bibfnamefont {T.}~\bibnamefont {Wichmann}},\ }\bibfield
  {title} {\enquote {\bibinfo {title} {Basal ganglia activity patterns in
  parkinsonism and computational modeling of their downstream effects},}\
  }\href@noop {} {\bibfield  {journal} {\bibinfo  {journal} {Eur.
  J.~Neurosci.}\ }\textbf {\bibinfo {volume} {36}},\ \bibinfo {pages}
  {2213--2228} (\bibinfo {year} {2012})}\BibitemShut {NoStop}%
\bibitem [{\citenamefont {Dahlem}(2013)}]{DAH13}%
  \BibitemOpen
  \bibfield  {author} {\bibinfo {author} {\bibfnamefont {M.~A.}\ \bibnamefont
  {Dahlem}},\ }\href@noop {} {\enquote {\bibinfo {title} {Migraine generator
  network and spreading depression dynamics as neuromodulation targets in
  episodic migraine},}\ }\bibinfo {howpublished} {Chaos (accepted)} (\bibinfo
  {year} {2013})\BibitemShut {NoStop}%
\bibitem [{\citenamefont {Magis}\ and\ \citenamefont {Schoenen}(2012)}]{MAG12}%
  \BibitemOpen
  \bibfield  {author} {\bibinfo {author} {\bibfnamefont {D.}~\bibnamefont
  {Magis}}\ and\ \bibinfo {author} {\bibfnamefont {J.}~\bibnamefont
  {Schoenen}},\ }\bibfield  {title} {\enquote {\bibinfo {title} {{{A}dvances
  and challenges in neurostimulation for headaches}},}\ }\href@noop {}
  {\bibfield  {journal} {\bibinfo  {journal} {Lancet Neurol.}\ }\textbf
  {\bibinfo {volume} {11}},\ \bibinfo {pages} {708--719} (\bibinfo {year}
  {2012})}\BibitemShut {NoStop}%
\bibitem [{\citenamefont {Koehler}\ and\ \citenamefont {Boes}(2010)}]{KOE10}%
  \BibitemOpen
  \bibfield  {author} {\bibinfo {author} {\bibfnamefont {P.~J.}\ \bibnamefont
  {Koehler}}\ and\ \bibinfo {author} {\bibfnamefont {C.~J.}\ \bibnamefont
  {Boes}},\ }\bibfield  {title} {\enquote {\bibinfo {title} {{{A} history of
  non-drug treatment in headache, particularly migraine}},}\ }\href@noop {}
  {\bibfield  {journal} {\bibinfo  {journal} {Brain}\ }\textbf {\bibinfo
  {volume} {133}},\ \bibinfo {pages} {2489--2500} (\bibinfo {year}
  {2010})}\BibitemShut {NoStop}%
\bibitem [{\citenamefont {Wiener}(1948)}]{WIE48}%
  \BibitemOpen
  \bibfield  {author} {\bibinfo {author} {\bibfnamefont {N.}~\bibnamefont
  {Wiener}},\ }\bibfield  {title} {\enquote {\bibinfo {title} {Cybernetics; or
  control and communication in the animal and the machine.}}\ }\href@noop {} {\
   (\bibinfo {year} {1948})}\BibitemShut {NoStop}%
\bibitem [{\citenamefont {Schiff}(2011)}]{SCH11e}%
  \BibitemOpen
  \bibfield  {author} {\bibinfo {author} {\bibfnamefont {S.~J.}\ \bibnamefont
  {Schiff}},\ }\href@noop {} {\emph {\bibinfo {title} {Neural Control
  Engineering: The Emerging Intersection Between Control Theory and
  Neuroscience}}}\ (\bibinfo  {publisher} {MIT Press},\ \bibinfo {address}
  {Cambridge, MA},\ \bibinfo {year} {2011})\BibitemShut {NoStop}%
\bibitem [{\citenamefont {Goodfellow}, \citenamefont {Schindler},\ and\
  \citenamefont {Baier}(2012)}]{GOO12}%
  \BibitemOpen
  \bibfield  {author} {\bibinfo {author} {\bibfnamefont {M.}~\bibnamefont
  {Goodfellow}}, \bibinfo {author} {\bibfnamefont {K.}~\bibnamefont
  {Schindler}}, \ and\ \bibinfo {author} {\bibfnamefont {G.}~\bibnamefont
  {Baier}},\ }\bibfield  {title} {\enquote {\bibinfo {title} {{{S}elf-organised
  transients in a neural mass model of epileptogenic tissue dynamics}},}\
  }\href@noop {} {\bibfield  {journal} {\bibinfo  {journal} {Neuroimage}\
  }\textbf {\bibinfo {volume} {59}},\ \bibinfo {pages} {2644--2660} (\bibinfo
  {year} {2012})}\BibitemShut {NoStop}%
\bibitem [{\citenamefont {Suffczynski}\ \emph {et~al.}(2006)\citenamefont
  {Suffczynski}, \citenamefont {Lopes~da Silva}, \citenamefont {Parra},
  \citenamefont {Velis}, \citenamefont {Bouwman}, \citenamefont {van Rijn},
  \citenamefont {van Hese}, \citenamefont {Boon}, \citenamefont {Khosravani},
  \citenamefont {Derchansky}, \citenamefont {Carlen},\ and\ \citenamefont
  {Kalitzin}}]{SUF06}%
  \BibitemOpen
  \bibfield  {author} {\bibinfo {author} {\bibfnamefont {P.}~\bibnamefont
  {Suffczynski}}, \bibinfo {author} {\bibfnamefont {F.~H.}\ \bibnamefont
  {Lopes~da Silva}}, \bibinfo {author} {\bibfnamefont {J.}~\bibnamefont
  {Parra}}, \bibinfo {author} {\bibfnamefont {D.~N.}\ \bibnamefont {Velis}},
  \bibinfo {author} {\bibfnamefont {B.~M.}\ \bibnamefont {Bouwman}}, \bibinfo
  {author} {\bibfnamefont {C.~M.}\ \bibnamefont {van Rijn}}, \bibinfo {author}
  {\bibfnamefont {P.}~\bibnamefont {van Hese}}, \bibinfo {author}
  {\bibfnamefont {P.}~\bibnamefont {Boon}}, \bibinfo {author} {\bibfnamefont
  {H.}~\bibnamefont {Khosravani}}, \bibinfo {author} {\bibfnamefont
  {M.}~\bibnamefont {Derchansky}}, \bibinfo {author} {\bibfnamefont
  {P.}~\bibnamefont {Carlen}}, \ and\ \bibinfo {author} {\bibfnamefont
  {S.}~\bibnamefont {Kalitzin}},\ }\bibfield  {title} {\enquote {\bibinfo
  {title} {{{D}ynamics of epileptic phenomena determined from statistics of
  ictal transitions}},}\ }\href@noop {} {\bibfield  {journal} {\bibinfo
  {journal} {IEEE Trans. Biomed. Eng.}\ }\textbf {\bibinfo {volume} {53}},\
  \bibinfo {pages} {524--532} (\bibinfo {year} {2006})}\BibitemShut {NoStop}%
\bibitem [{\citenamefont {Giffin}\ \emph {et~al.}(2003)\citenamefont {Giffin},
  \citenamefont {Ruggiero}, \citenamefont {Lipton}, \citenamefont
  {Silberstein}, \citenamefont {Tvedskov}, \citenamefont {Olesen},
  \citenamefont {Altman}, \citenamefont {Goadsby},\ and\ \citenamefont
  {Macrae}}]{GIF03}%
  \BibitemOpen
  \bibfield  {author} {\bibinfo {author} {\bibfnamefont {N.~J.}\ \bibnamefont
  {Giffin}}, \bibinfo {author} {\bibfnamefont {L.}~\bibnamefont {Ruggiero}},
  \bibinfo {author} {\bibfnamefont {R.~B.}\ \bibnamefont {Lipton}}, \bibinfo
  {author} {\bibfnamefont {S.~D.}\ \bibnamefont {Silberstein}}, \bibinfo
  {author} {\bibfnamefont {J.~F.}\ \bibnamefont {Tvedskov}}, \bibinfo {author}
  {\bibfnamefont {J.}~\bibnamefont {Olesen}}, \bibinfo {author} {\bibfnamefont
  {J.}~\bibnamefont {Altman}}, \bibinfo {author} {\bibfnamefont {P.~J.}\
  \bibnamefont {Goadsby}}, \ and\ \bibinfo {author} {\bibfnamefont
  {A.}~\bibnamefont {Macrae}},\ }\bibfield  {title} {\enquote {\bibinfo {title}
  {{{P}remonitory symptoms in migraine: an electronic diary study}},}\
  }\href@noop {} {\bibfield  {journal} {\bibinfo  {journal} {Neurology}\
  }\textbf {\bibinfo {volume} {60}},\ \bibinfo {pages} {935--940} (\bibinfo
  {year} {2003})}\BibitemShut {NoStop}%
\bibitem [{\citenamefont {Charles}(2012)}]{CHA12}%
  \BibitemOpen
  \bibfield  {author} {\bibinfo {author} {\bibfnamefont {A.}~\bibnamefont
  {Charles}},\ }\bibfield  {title} {\enquote {\bibinfo {title} {{{M}igraine is
  not primarily a vascular disorder}},}\ }\href@noop {} {\bibfield  {journal}
  {\bibinfo  {journal} {Cephalalgia}\ }\textbf {\bibinfo {volume} {32}},\
  \bibinfo {pages} {431--432} (\bibinfo {year} {2012})}\BibitemShut {NoStop}%
\bibitem [{\citenamefont {Olesen}(2013)}]{OLE13}%
  \BibitemOpen
  \bibfield  {author} {\bibinfo {author} {\bibfnamefont {J.}~\bibnamefont
  {Olesen}},\ }\bibfield  {title} {\enquote {\bibinfo {title} {{{T}he
  {I}nternational {C}lassification of {H}eadache {D}isorders, 3rd edition (beta
  version)}},}\ }\href@noop {} {\bibfield  {journal} {\bibinfo  {journal}
  {Cephalalgia}\ }\textbf {\bibinfo {volume} {33}},\ \bibinfo {pages}
  {629--808} (\bibinfo {year} {2013})}\BibitemShut {NoStop}%
\bibitem [{\citenamefont {Vincent}\ and\ \citenamefont
  {Hadjikhani}(2007)}]{VIN07}%
  \BibitemOpen
  \bibfield  {author} {\bibinfo {author} {\bibfnamefont {M.}~\bibnamefont
  {Vincent}}\ and\ \bibinfo {author} {\bibfnamefont {N.}~\bibnamefont
  {Hadjikhani}},\ }\bibfield  {title} {\enquote {\bibinfo {title} {{{M}igraine
  aura and related phenomena: beyond scotomata and scintillations}},}\
  }\href@noop {} {\bibfield  {journal} {\bibinfo  {journal} {Cephalalgia}\
  }\textbf {\bibinfo {volume} {27}},\ \bibinfo {pages} {1368--1377} (\bibinfo
  {year} {2007})}\BibitemShut {NoStop}%
\bibitem [{\citenamefont {Karatas}\ \emph {et~al.}(2013)\citenamefont
  {Karatas}, \citenamefont {Erdener}, \citenamefont {Gursoy-Ozdemir},
  \citenamefont {Lule}, \citenamefont {Eren-Kocak}, \citenamefont {Sen},\ and\
  \citenamefont {Dalkara}}]{KAR13}%
  \BibitemOpen
  \bibfield  {author} {\bibinfo {author} {\bibfnamefont {H.}~\bibnamefont
  {Karatas}}, \bibinfo {author} {\bibfnamefont {S.~E.}\ \bibnamefont
  {Erdener}}, \bibinfo {author} {\bibfnamefont {Y.}~\bibnamefont
  {Gursoy-Ozdemir}}, \bibinfo {author} {\bibfnamefont {S.}~\bibnamefont
  {Lule}}, \bibinfo {author} {\bibfnamefont {E.}~\bibnamefont {Eren-Kocak}},
  \bibinfo {author} {\bibfnamefont {Z.~D.}\ \bibnamefont {Sen}}, \ and\
  \bibinfo {author} {\bibfnamefont {T.}~\bibnamefont {Dalkara}},\ }\bibfield
  {title} {\enquote {\bibinfo {title} {{{S}preading depression triggers
  headache by activating neuronal {P}anx1 channels}},}\ }\href@noop {}
  {\bibfield  {journal} {\bibinfo  {journal} {Science}\ }\textbf {\bibinfo
  {volume} {339}},\ \bibinfo {pages} {1092--1095} (\bibinfo {year}
  {2013})}\BibitemShut {NoStop}%
\bibitem [{\citenamefont {Rasmussen}\ and\ \citenamefont
  {Olesen}(1992)}]{RAS92}%
  \BibitemOpen
  \bibfield  {author} {\bibinfo {author} {\bibfnamefont {B.~K.}\ \bibnamefont
  {Rasmussen}}\ and\ \bibinfo {author} {\bibfnamefont {J.}~\bibnamefont
  {Olesen}},\ }\bibfield  {title} {\enquote {\bibinfo {title} {{{M}igraine with
  aura and migraine without aura: an epidemiological study}},}\ }\href@noop {}
  {\bibfield  {journal} {\bibinfo  {journal} {Cephalalgia}\ }\textbf {\bibinfo
  {volume} {12}},\ \bibinfo {pages} {221--228} (\bibinfo {year}
  {1992})}\BibitemShut {NoStop}%
\bibitem [{\citenamefont {Ahn}(2010)}]{AHN10}%
  \BibitemOpen
  \bibfield  {author} {\bibinfo {author} {\bibfnamefont {A.~H.}\ \bibnamefont
  {Ahn}},\ }\bibfield  {title} {\enquote {\bibinfo {title} {{{O}n the temporal
  relationship between throbbing migraine pain and arterial pulse}},}\
  }\href@noop {} {\bibfield  {journal} {\bibinfo  {journal} {Headache}\
  }\textbf {\bibinfo {volume} {50}},\ \bibinfo {pages} {1507--1510} (\bibinfo
  {year} {2010})}\BibitemShut {NoStop}%
\bibitem [{\citenamefont {Mo}\ \emph {et~al.}(2013)\citenamefont {Mo},
  \citenamefont {Maizels}, \citenamefont {Ding},\ and\ \citenamefont
  {Ahn}}]{MO13}%
  \BibitemOpen
  \bibfield  {author} {\bibinfo {author} {\bibfnamefont {J.}~\bibnamefont
  {Mo}}, \bibinfo {author} {\bibfnamefont {M.}~\bibnamefont {Maizels}},
  \bibinfo {author} {\bibfnamefont {M.}~\bibnamefont {Ding}}, \ and\ \bibinfo
  {author} {\bibfnamefont {A.~H.}\ \bibnamefont {Ahn}},\ }\bibfield  {title}
  {\enquote {\bibinfo {title} {{{D}oes throbbing pain have a brain
  signature?}}}\ }\href@noop {} {\bibfield  {journal} {\bibinfo  {journal}
  {Pain}\ }\textbf {\bibinfo {volume} {154}},\ \bibinfo {pages} {1150--1155}
  (\bibinfo {year} {2013})}\BibitemShut {NoStop}%
\bibitem [{\citenamefont {Brandes}(2008)}]{BRA08}%
  \BibitemOpen
  \bibfield  {author} {\bibinfo {author} {\bibfnamefont {J.~L.}\ \bibnamefont
  {Brandes}},\ }\bibfield  {title} {\enquote {\bibinfo {title} {{{T}he migraine
  cycle: patient burden of migraine during and between migraine attacks}},}\
  }\href@noop {} {\bibfield  {journal} {\bibinfo  {journal} {Headache}\
  }\textbf {\bibinfo {volume} {48}},\ \bibinfo {pages} {430--441} (\bibinfo
  {year} {2008})}\BibitemShut {NoStop}%
\bibitem [{\citenamefont {Weiller}\ \emph {et~al.}(1995)\citenamefont
  {Weiller}, \citenamefont {May}, \citenamefont {Limmroth}, \citenamefont
  {Juptner}, \citenamefont {Kaube}, \citenamefont {Schayck}, \citenamefont
  {Coenen},\ and\ \citenamefont {Diener}}]{WEI95a}%
  \BibitemOpen
  \bibfield  {author} {\bibinfo {author} {\bibfnamefont {C.}~\bibnamefont
  {Weiller}}, \bibinfo {author} {\bibfnamefont {A.}~\bibnamefont {May}},
  \bibinfo {author} {\bibfnamefont {V.}~\bibnamefont {Limmroth}}, \bibinfo
  {author} {\bibfnamefont {M.}~\bibnamefont {Juptner}}, \bibinfo {author}
  {\bibfnamefont {H.}~\bibnamefont {Kaube}}, \bibinfo {author} {\bibfnamefont
  {R.~V.}\ \bibnamefont {Schayck}}, \bibinfo {author} {\bibfnamefont {H.~H.}\
  \bibnamefont {Coenen}}, \ and\ \bibinfo {author} {\bibfnamefont {H.~C.}\
  \bibnamefont {Diener}},\ }\bibfield  {title} {\enquote {\bibinfo {title}
  {{{B}rain stem activation in spontaneous human migraine attacks}},}\
  }\href@noop {} {\bibfield  {journal} {\bibinfo  {journal} {Nat. Med.}\
  }\textbf {\bibinfo {volume} {1}},\ \bibinfo {pages} {658--660} (\bibinfo
  {year} {1995})}\BibitemShut {NoStop}%
\bibitem [{\citenamefont {Welch}\ \emph {et~al.}(2001)\citenamefont {Welch},
  \citenamefont {Nagesh}, \citenamefont {Aurora},\ and\ \citenamefont
  {Gelman}}]{WEL01}%
  \BibitemOpen
  \bibfield  {author} {\bibinfo {author} {\bibfnamefont {K.}~\bibnamefont
  {Welch}}, \bibinfo {author} {\bibfnamefont {V.}~\bibnamefont {Nagesh}},
  \bibinfo {author} {\bibfnamefont {S.~K.}\ \bibnamefont {Aurora}}, \ and\
  \bibinfo {author} {\bibfnamefont {N.}~\bibnamefont {Gelman}},\ }\bibfield
  {title} {\enquote {\bibinfo {title} {Periaqueductal gray matter dysfunction
  in migraine: cause or the burden of illness?}}\ }\href@noop {} {\bibfield
  {journal} {\bibinfo  {journal} {Headache: The Journal of Head and Face Pain}\
  }\textbf {\bibinfo {volume} {41}},\ \bibinfo {pages} {629--637} (\bibinfo
  {year} {2001})}\BibitemShut {NoStop}%
\bibitem [{\citenamefont {Olesen}, \citenamefont {Larsen},\ and\ \citenamefont
  {Lauritzen}(1981)}]{OLE81}%
  \BibitemOpen
  \bibfield  {author} {\bibinfo {author} {\bibfnamefont {J.}~\bibnamefont
  {Olesen}}, \bibinfo {author} {\bibfnamefont {B.}~\bibnamefont {Larsen}}, \
  and\ \bibinfo {author} {\bibfnamefont {M.}~\bibnamefont {Lauritzen}},\
  }\bibfield  {title} {\enquote {\bibinfo {title} {{{F}ocal hyperemia followed
  by spreading oligemia and impaired activation of r{C}{B}{F} in classic
  migraine}},}\ }\href@noop {} {\bibfield  {journal} {\bibinfo  {journal} {Ann.
  Neurol.}\ }\textbf {\bibinfo {volume} {9}},\ \bibinfo {pages} {344--352}
  (\bibinfo {year} {1981})}\BibitemShut {NoStop}%
\bibitem [{\citenamefont {Hadjikhani}\ \emph {et~al.}(2001)\citenamefont
  {Hadjikhani}, \citenamefont {Sanchez Del.~Rio.}, \citenamefont {Wu},
  \citenamefont {Schwartz}, \citenamefont {Bakker}, \citenamefont {Fischl},
  \citenamefont {Kwong}, \citenamefont {Cutrer}, \citenamefont {Rosen},
  \citenamefont {Tootell}, \citenamefont {Sorensen},\ and\ \citenamefont
  {Moskowitz}}]{HAD01}%
  \BibitemOpen
  \bibfield  {author} {\bibinfo {author} {\bibfnamefont {N.}~\bibnamefont
  {Hadjikhani}}, \bibinfo {author} {\bibfnamefont {M.}~\bibnamefont {Sanchez
  Del.~Rio.}}, \bibinfo {author} {\bibfnamefont {O.}~\bibnamefont {Wu}},
  \bibinfo {author} {\bibfnamefont {D.}~\bibnamefont {Schwartz}}, \bibinfo
  {author} {\bibfnamefont {D.}~\bibnamefont {Bakker}}, \bibinfo {author}
  {\bibfnamefont {B.}~\bibnamefont {Fischl}}, \bibinfo {author} {\bibfnamefont
  {K.~K.}\ \bibnamefont {Kwong}}, \bibinfo {author} {\bibfnamefont {F.~M.}\
  \bibnamefont {Cutrer}}, \bibinfo {author} {\bibfnamefont {B.~R.}\
  \bibnamefont {Rosen}}, \bibinfo {author} {\bibfnamefont {R.~B.}\ \bibnamefont
  {Tootell}}, \bibinfo {author} {\bibfnamefont {A.~G.}\ \bibnamefont
  {Sorensen}}, \ and\ \bibinfo {author} {\bibfnamefont {M.~A.}\ \bibnamefont
  {Moskowitz}},\ }\bibfield  {title} {\enquote {\bibinfo {title} {{Mechanisms
  of migraine aura revealed by functional MRI in human visual cortex}},}\
  }\href@noop {} {\bibfield  {journal} {\bibinfo  {journal} {Proc. Natl. Acad.
  Sci. U.S.A.}\ }\textbf {\bibinfo {volume} {98}},\ \bibinfo {pages}
  {4687--4692} (\bibinfo {year} {2001})}\BibitemShut {NoStop}%
\bibitem [{\citenamefont {Lauritzen}(1994)}]{LAU94}%
  \BibitemOpen
  \bibfield  {author} {\bibinfo {author} {\bibfnamefont {M.}~\bibnamefont
  {Lauritzen}},\ }\bibfield  {title} {\enquote {\bibinfo {title}
  {Pathophysiology of the migraine aura. {T}he spreading depression theory},}\
  }\href@noop {} {\bibfield  {journal} {\bibinfo  {journal} {Brain}\ }\textbf
  {\bibinfo {volume} {117}},\ \bibinfo {pages} {199--210} (\bibinfo {year}
  {1994})}\BibitemShut {NoStop}%
\bibitem [{\citenamefont {Hansen}\ \emph {et~al.}(2012)\citenamefont {Hansen},
  \citenamefont {Lipton}, \citenamefont {Dodick}, \citenamefont {Silberstein},
  \citenamefont {Saper}, \citenamefont {Aurora}, \citenamefont {Goadsby},\ and\
  \citenamefont {Charles}}]{HAN12a}%
  \BibitemOpen
  \bibfield  {author} {\bibinfo {author} {\bibfnamefont {J.~M.}\ \bibnamefont
  {Hansen}}, \bibinfo {author} {\bibfnamefont {R.~B.}\ \bibnamefont {Lipton}},
  \bibinfo {author} {\bibfnamefont {D.~W.}\ \bibnamefont {Dodick}}, \bibinfo
  {author} {\bibfnamefont {S.~D.}\ \bibnamefont {Silberstein}}, \bibinfo
  {author} {\bibfnamefont {J.~R.}\ \bibnamefont {Saper}}, \bibinfo {author}
  {\bibfnamefont {S.~K.}\ \bibnamefont {Aurora}}, \bibinfo {author}
  {\bibfnamefont {P.~J.}\ \bibnamefont {Goadsby}}, \ and\ \bibinfo {author}
  {\bibfnamefont {A.}~\bibnamefont {Charles}},\ }\bibfield  {title} {\enquote
  {\bibinfo {title} {{{M}igraine headache is present in the aura phase: a
  prospective study}},}\ }\href@noop {} {\bibfield  {journal} {\bibinfo
  {journal} {Neurology}\ }\textbf {\bibinfo {volume} {79}},\ \bibinfo {pages}
  {2044--2049} (\bibinfo {year} {2012})}\BibitemShut {NoStop}%
\bibitem [{\citenamefont {Noseda}\ \emph {et~al.}(2010)\citenamefont {Noseda},
  \citenamefont {Kainz}, \citenamefont {Jakubowski}, \citenamefont {Gooley},
  \citenamefont {Saper}, \citenamefont {Digre},\ and\ \citenamefont
  {Burstein}}]{NOS10}%
  \BibitemOpen
  \bibfield  {author} {\bibinfo {author} {\bibfnamefont {R.}~\bibnamefont
  {Noseda}}, \bibinfo {author} {\bibfnamefont {V.}~\bibnamefont {Kainz}},
  \bibinfo {author} {\bibfnamefont {M.}~\bibnamefont {Jakubowski}}, \bibinfo
  {author} {\bibfnamefont {J.~J.}\ \bibnamefont {Gooley}}, \bibinfo {author}
  {\bibfnamefont {C.~B.}\ \bibnamefont {Saper}}, \bibinfo {author}
  {\bibfnamefont {K.}~\bibnamefont {Digre}}, \ and\ \bibinfo {author}
  {\bibfnamefont {R.}~\bibnamefont {Burstein}},\ }\bibfield  {title} {\enquote
  {\bibinfo {title} {A neural mechanism for exacerbation of headache by
  light},}\ }\href@noop {} {\bibfield  {journal} {\bibinfo  {journal} {Nature
  Neurosci.}\ }\textbf {\bibinfo {volume} {13}},\ \bibinfo {pages} {239--245}
  (\bibinfo {year} {2010})}\BibitemShut {NoStop}%
\bibitem [{\citenamefont {Summ}\ \emph {et~al.}(2010)\citenamefont {Summ},
  \citenamefont {Charbit}, \citenamefont {Andreou},\ and\ \citenamefont
  {Goadsby}}]{SUM10}%
  \BibitemOpen
  \bibfield  {author} {\bibinfo {author} {\bibfnamefont {O.}~\bibnamefont
  {Summ}}, \bibinfo {author} {\bibfnamefont {A.~R.}\ \bibnamefont {Charbit}},
  \bibinfo {author} {\bibfnamefont {A.~P.}\ \bibnamefont {Andreou}}, \ and\
  \bibinfo {author} {\bibfnamefont {P.~J.}\ \bibnamefont {Goadsby}},\
  }\bibfield  {title} {\enquote {\bibinfo {title} {{{M}odulation of
  nocioceptive transmission with calcitonin gene-related peptide receptor
  antagonists in the thalamus}},}\ }\href@noop {} {\bibfield  {journal}
  {\bibinfo  {journal} {Brain}\ }\textbf {\bibinfo {volume} {133}},\ \bibinfo
  {pages} {2540--2548} (\bibinfo {year} {2010})}\BibitemShut {NoStop}%
\bibitem [{\citenamefont {Reshodko}\ and\ \citenamefont {Bures}(1975)}]{RES75}%
  \BibitemOpen
  \bibfield  {author} {\bibinfo {author} {\bibfnamefont {L.~V.}\ \bibnamefont
  {Reshodko}}\ and\ \bibinfo {author} {\bibfnamefont {J.}~\bibnamefont
  {Bures}},\ }\bibfield  {title} {\enquote {\bibinfo {title} {{{C}omputer
  simulation of reverberating spreading depression in a network of cell
  automata}},}\ }\href@noop {} {\bibfield  {journal} {\bibinfo  {journal}
  {Biol. Cybern.}\ }\textbf {\bibinfo {volume} {18}},\ \bibinfo {pages}
  {181--189} (\bibinfo {year} {1975})}\BibitemShut {NoStop}%
\bibitem [{\citenamefont {Dahlem}\ and\ \citenamefont
  {M{\"u}ller}(1997)}]{DAH97}%
  \BibitemOpen
  \bibfield  {author} {\bibinfo {author} {\bibfnamefont {M.~A.}\ \bibnamefont
  {Dahlem}}\ and\ \bibinfo {author} {\bibfnamefont {S.~C.}\ \bibnamefont
  {M{\"u}ller}},\ }\bibfield  {title} {\enquote {\bibinfo {title} {Self-induced
  splitting of spiral-shaped spreading depression waves in chicken retina},}\
  }\href@noop {} {\bibfield  {journal} {\bibinfo  {journal} {Exp. Brain Res.}\
  }\textbf {\bibinfo {volume} {115}},\ \bibinfo {pages} {319--324} (\bibinfo
  {year} {1997})}\BibitemShut {NoStop}%
\bibitem [{\citenamefont {Tuckwell}\ and\ \citenamefont {Miura}(1978)}]{TUC78}%
  \BibitemOpen
  \bibfield  {author} {\bibinfo {author} {\bibfnamefont {H.~C.}\ \bibnamefont
  {Tuckwell}}\ and\ \bibinfo {author} {\bibfnamefont {R.~M.}\ \bibnamefont
  {Miura}},\ }\bibfield  {title} {\enquote {\bibinfo {title} {A mathematical
  model for spreading cortical depression},}\ }\href@noop {} {\bibfield
  {journal} {\bibinfo  {journal} {Biophys.~J.}\ }\textbf {\bibinfo {volume}
  {23}},\ \bibinfo {pages} {257--276} (\bibinfo {year} {1978})}\BibitemShut
  {NoStop}%
\bibitem [{\citenamefont {Miura}, \citenamefont {Huang},\ and\ \citenamefont
  {Wylie}(2007)}]{MIU07}%
  \BibitemOpen
  \bibfield  {author} {\bibinfo {author} {\bibfnamefont {R.~M.}\ \bibnamefont
  {Miura}}, \bibinfo {author} {\bibfnamefont {H.}~\bibnamefont {Huang}}, \ and\
  \bibinfo {author} {\bibfnamefont {J.~J.}\ \bibnamefont {Wylie}},\ }\bibfield
  {title} {\enquote {\bibinfo {title} {Cortical spreading depression: An
  enigma},}\ }\href@noop {} {\bibfield  {journal} {\bibinfo  {journal} {Eur.
  Phys. J. Spec. Top.}\ }\textbf {\bibinfo {volume} {147}},\ \bibinfo {pages}
  {287--302} (\bibinfo {year} {2007})}\BibitemShut {NoStop}%
\bibitem [{\citenamefont {Somjen}(2001)}]{SOM01}%
  \BibitemOpen
  \bibfield  {author} {\bibinfo {author} {\bibfnamefont {G.~G.}\ \bibnamefont
  {Somjen}},\ }\bibfield  {title} {\enquote {\bibinfo {title} {Mechanisms of
  spreading depression and hypoxic spreading depression-like depolarization},}\
  }\href@noop {} {\bibfield  {journal} {\bibinfo  {journal} {Physiol. Rev.}\
  }\textbf {\bibinfo {volume} {81}},\ \bibinfo {pages} {1065--1096} (\bibinfo
  {year} {2001})}\BibitemShut {NoStop}%
\bibitem [{\citenamefont {Ayata}(2013)}]{AYA13}%
  \BibitemOpen
  \bibfield  {author} {\bibinfo {author} {\bibfnamefont {C.}~\bibnamefont
  {Ayata}},\ }\bibfield  {title} {\enquote {\bibinfo {title} {{{S}preading
  depression and neurovascular coupling}},}\ }\href@noop {} {\bibfield
  {journal} {\bibinfo  {journal} {Stroke}\ }\textbf {\bibinfo {volume} {44}},\
  \bibinfo {pages} {S87--89} (\bibinfo {year} {2013})}\BibitemShut {NoStop}%
\bibitem [{\citenamefont {Chang}\ \emph {et~al.}(2012)\citenamefont {Chang},
  \citenamefont {Brennan}, \citenamefont {He}, \citenamefont {Huang},
  \citenamefont {Miura}, \citenamefont {Wilson},\ and\ \citenamefont
  {Wylie}}]{CHA12a}%
  \BibitemOpen
  \bibfield  {author} {\bibinfo {author} {\bibfnamefont {J.~C.}\ \bibnamefont
  {Chang}}, \bibinfo {author} {\bibfnamefont {K.}~\bibnamefont {Brennan}},
  \bibinfo {author} {\bibfnamefont {D.}~\bibnamefont {He}}, \bibinfo {author}
  {\bibfnamefont {H.}~\bibnamefont {Huang}}, \bibinfo {author} {\bibfnamefont
  {R.~M.}\ \bibnamefont {Miura}}, \bibinfo {author} {\bibfnamefont {P.~L.}\
  \bibnamefont {Wilson}}, \ and\ \bibinfo {author} {\bibfnamefont {J.~J.}\
  \bibnamefont {Wylie}},\ }\bibfield  {title} {\enquote {\bibinfo {title} {A
  mathematical model of the metabolic and perfusion effects on cortical
  spreading depression},}\ }\href@noop {} {\bibfield  {journal} {\bibinfo
  {journal} {arXiv preprint arXiv:1207.3563}\ } (\bibinfo {year}
  {2012})}\BibitemShut {NoStop}%
\bibitem [{\citenamefont {Hodgkin}\ and\ \citenamefont {Huxley}(1952)}]{HOD52}%
  \BibitemOpen
  \bibfield  {author} {\bibinfo {author} {\bibfnamefont {A.~L.}\ \bibnamefont
  {Hodgkin}}\ and\ \bibinfo {author} {\bibfnamefont {A.~F.}\ \bibnamefont
  {Huxley}},\ }\bibfield  {title} {\enquote {\bibinfo {title} {A quantitative
  description of membrane current and its application to conduction and
  excitation in nerve},}\ }\href@noop {} {\bibfield  {journal} {\bibinfo
  {journal} {J.~Physiol.}\ }\textbf {\bibinfo {volume} {117}},\ \bibinfo
  {pages} {500} (\bibinfo {year} {1952})}\BibitemShut {NoStop}%
\bibitem [{\citenamefont {Dreier}\ \emph {et~al.}(2013)\citenamefont {Dreier},
  \citenamefont {Isele}, \citenamefont {Reiffurth}, \citenamefont {Kirov},
  \citenamefont {Dahlem},\ and\ \citenamefont {Herreras}}]{DRE12}%
  \BibitemOpen
  \bibfield  {author} {\bibinfo {author} {\bibfnamefont {J.~P.}\ \bibnamefont
  {Dreier}}, \bibinfo {author} {\bibfnamefont {T.~M.}\ \bibnamefont {Isele}},
  \bibinfo {author} {\bibfnamefont {C.}~\bibnamefont {Reiffurth}}, \bibinfo
  {author} {\bibfnamefont {S.~A.}\ \bibnamefont {Kirov}}, \bibinfo {author}
  {\bibfnamefont {M.~A.}\ \bibnamefont {Dahlem}}, \ and\ \bibinfo {author}
  {\bibfnamefont {O.}~\bibnamefont {Herreras}},\ }\bibfield  {title} {\enquote
  {\bibinfo {title} {Is spreading depolarization characterized by an abrupt,
  massive release of {G}ibbs free energy from the human brain cortex?}}\ }\href
  {\doibase 10.1177/1073858412453340} {\bibfield  {journal} {\bibinfo
  {journal} {Neuroscientist}\ }\textbf {\bibinfo {volume} {19}},\ \bibinfo
  {pages} {25--42} (\bibinfo {year} {2013})}\BibitemShut {NoStop}%
\bibitem [{\citenamefont {Kager}, \citenamefont {Wadman},\ and\ \citenamefont
  {Somjen}(2000)}]{KAG00}%
  \BibitemOpen
  \bibfield  {author} {\bibinfo {author} {\bibfnamefont {H.}~\bibnamefont
  {Kager}}, \bibinfo {author} {\bibfnamefont {W.~J.}\ \bibnamefont {Wadman}}, \
  and\ \bibinfo {author} {\bibfnamefont {G.~G.}\ \bibnamefont {Somjen}},\
  }\bibfield  {title} {\enquote {\bibinfo {title} {Simulated seizures and
  spreading depression in a neuron model incorporating interstitial space and
  ion concentrations},}\ }\href@noop {} {\bibfield  {journal} {\bibinfo
  {journal} {J.~Neurophysiol.}\ }\textbf {\bibinfo {volume} {84}},\ \bibinfo
  {pages} {495--512} (\bibinfo {year} {2000})}\BibitemShut {NoStop}%
\bibitem [{\citenamefont {Shapiro}(2001)}]{SHA01}%
  \BibitemOpen
  \bibfield  {author} {\bibinfo {author} {\bibfnamefont {B.~E.}\ \bibnamefont
  {Shapiro}},\ }\bibfield  {title} {\enquote {\bibinfo {title} {Osmotic forces
  and gap junctions in spreading depression: a computational model},}\
  }\href@noop {} {\bibfield  {journal} {\bibinfo  {journal} {J. Comput.
  Neurosci.}\ }\textbf {\bibinfo {volume} {10}},\ \bibinfo {pages} {99--120}
  (\bibinfo {year} {2001})}\BibitemShut {NoStop}%
\bibitem [{\citenamefont {Somjen}, \citenamefont {Kager},\ and\ \citenamefont
  {Wadman}(2008)}]{SOM08}%
  \BibitemOpen
  \bibfield  {author} {\bibinfo {author} {\bibfnamefont {G.~G.}\ \bibnamefont
  {Somjen}}, \bibinfo {author} {\bibfnamefont {H.}~\bibnamefont {Kager}}, \
  and\ \bibinfo {author} {\bibfnamefont {W.~J.}\ \bibnamefont {Wadman}},\
  }\bibfield  {title} {\enquote {\bibinfo {title} {Computer simulations of
  neuron-glia interactions mediated by ion flux},}\ }\href@noop {} {\bibfield
  {journal} {\bibinfo  {journal} {J. Comput. Neurosci.}\ }\textbf {\bibinfo
  {volume} {25}},\ \bibinfo {pages} {349--365} (\bibinfo {year}
  {2008})}\BibitemShut {NoStop}%
\bibitem [{\citenamefont {Yao}, \citenamefont {Huang},\ and\ \citenamefont
  {Miura}(2011)}]{YAO11a}%
  \BibitemOpen
  \bibfield  {author} {\bibinfo {author} {\bibfnamefont {W.}~\bibnamefont
  {Yao}}, \bibinfo {author} {\bibfnamefont {H.}~\bibnamefont {Huang}}, \ and\
  \bibinfo {author} {\bibfnamefont {R.~M.}\ \bibnamefont {Miura}},\ }\bibfield
  {title} {\enquote {\bibinfo {title} {{{A} continuum neuronal model for the
  instigation and propagation of cortical spreading depression}},}\ }\href@noop
  {} {\bibfield  {journal} {\bibinfo  {journal} {Bull. Math. Biol.}\ }\textbf
  {\bibinfo {volume} {73}},\ \bibinfo {pages} {2773--2790} (\bibinfo {year}
  {2011})}\BibitemShut {NoStop}%
\bibitem [{\citenamefont {Bressloff}(2012)}]{BRE12}%
  \BibitemOpen
  \bibfield  {author} {\bibinfo {author} {\bibfnamefont {P.~C.}\ \bibnamefont
  {Bressloff}},\ }\bibfield  {title} {\enquote {\bibinfo {title}
  {Spatiotemporal dynamics of continuum neural fields},}\ }\href@noop {}
  {\bibfield  {journal} {\bibinfo  {journal} {J.~Phys.~A}\ }\textbf {\bibinfo
  {volume} {45}},\ \bibinfo {pages} {033001} (\bibinfo {year}
  {2012})}\BibitemShut {NoStop}%
\bibitem [{\citenamefont {Ullah}\ \emph {et~al.}(2009)\citenamefont {Ullah},
  \citenamefont {Cressman~Jr.}, \citenamefont {Barreto},\ and\ \citenamefont
  {Schiff}}]{ULL08}%
  \BibitemOpen
  \bibfield  {author} {\bibinfo {author} {\bibfnamefont {G.}~\bibnamefont
  {Ullah}}, \bibinfo {author} {\bibfnamefont {J.~R.}\ \bibnamefont
  {Cressman~Jr.}}, \bibinfo {author} {\bibfnamefont {E.}~\bibnamefont
  {Barreto}}, \ and\ \bibinfo {author} {\bibfnamefont {S.~J.}\ \bibnamefont
  {Schiff}},\ }\bibfield  {title} {\enquote {\bibinfo {title} {The influence of
  sodium and potassium dynamics on excitability, seizures, and the stability of
  persistent states: Ii. network and glial dynamics},}\ }\href {\doibase
  10.1007/s10827-008-0130-6} {\bibfield  {journal} {\bibinfo  {journal}
  {J.~Comput. Neurosci.}\ }\textbf {\bibinfo {volume} {26}},\ \bibinfo {pages}
  {171--183} (\bibinfo {year} {2009})},\ \bibinfo {note}
  {arXiv:0806.3741v2}\BibitemShut {NoStop}%
\bibitem [{\citenamefont {Dahlem}\ and\ \citenamefont
  {Hadjikhani}(2009)}]{DAH08d}%
  \BibitemOpen
  \bibfield  {author} {\bibinfo {author} {\bibfnamefont {M.~A.}\ \bibnamefont
  {Dahlem}}\ and\ \bibinfo {author} {\bibfnamefont {N.}~\bibnamefont
  {Hadjikhani}},\ }\bibfield  {title} {\enquote {\bibinfo {title} {Migraine
  aura: retracting particle-like waves in weakly susceptible cortex},}\ }\href
  {\doibase doi:10.1371/journal.pone.0005007} {\bibfield  {journal} {\bibinfo
  {journal} {PLoS ONE}\ }\textbf {\bibinfo {volume} {4}},\ \bibinfo {pages}
  {e5007} (\bibinfo {year} {2009})}\BibitemShut {NoStop}%
\bibitem [{\citenamefont {Dahlem}\ and\ \citenamefont {Isele}(2013)}]{DAH12b}%
  \BibitemOpen
  \bibfield  {author} {\bibinfo {author} {\bibfnamefont {M.~A.}\ \bibnamefont
  {Dahlem}}\ and\ \bibinfo {author} {\bibfnamefont {T.~M.}\ \bibnamefont
  {Isele}},\ }\bibfield  {title} {\enquote {\bibinfo {title} {Transient
  localized wave patterns and their application to migraine},}\ }\href
  {\doibase 10.1186/2190-8567-3-7} {\bibfield  {journal} {\bibinfo  {journal}
  {J. Math. Neurosci}\ }\textbf {\bibinfo {volume} {3}},\ \bibinfo {pages} {7}
  (\bibinfo {year} {2013})}\BibitemShut {NoStop}%
\bibitem [{\citenamefont {Akhmediev}\ and\ \citenamefont
  {Ankiewicz}(2008)}]{AKH08}%
  \BibitemOpen
  \bibfield  {author} {\bibinfo {author} {\bibfnamefont {N.}~\bibnamefont
  {Akhmediev}}\ and\ \bibinfo {author} {\bibfnamefont {A.}~\bibnamefont
  {Ankiewicz}},\ }\href@noop {} {\emph {\bibinfo {title} {Dissipative
  {S}olitons: {F}rom {O}ptics to {B}iology and {M}edicine}}},\ edited by\
  \bibinfo {editor} {\bibfnamefont {N.}~\bibnamefont {Akhmediev}}\ and\
  \bibinfo {editor} {\bibfnamefont {A.}~\bibnamefont {Ankiewicz}},\ Vol.\
  \bibinfo {volume} {751}\ (\bibinfo  {publisher} {Springer},\ \bibinfo {year}
  {2008})\BibitemShut {NoStop}%
\bibitem [{\citenamefont {Kerner}\ and\ \citenamefont {Osipov}(1994)}]{KER94}%
  \BibitemOpen
  \bibfield  {author} {\bibinfo {author} {\bibfnamefont {B.~S.}\ \bibnamefont
  {Kerner}}\ and\ \bibinfo {author} {\bibfnamefont {V.~V.}\ \bibnamefont
  {Osipov}},\ }\href@noop {} {\emph {\bibinfo {title} {Autosolitons: {A} new
  approach to problems of self-organization and turbulence}}}\ (\bibinfo
  {publisher} {Kluwer Academic Publishers},\ \bibinfo {address} {Dordrecht},\
  \bibinfo {year} {1994})\BibitemShut {NoStop}%
\bibitem [{\citenamefont {Strong}\ \emph {et~al.}(2007)\citenamefont {Strong},
  \citenamefont {Anderson}, \citenamefont {Watts}, \citenamefont {Virley},
  \citenamefont {Lloyd}, \citenamefont {Irving}, \citenamefont {Nagafuji},
  \citenamefont {Ninomiya}, \citenamefont {Nakamura}, \citenamefont {Dunn},\
  and\ \citenamefont {Graf}}]{STR07}%
  \BibitemOpen
  \bibfield  {author} {\bibinfo {author} {\bibfnamefont {A.~J.}\ \bibnamefont
  {Strong}}, \bibinfo {author} {\bibfnamefont {P.~J.}\ \bibnamefont
  {Anderson}}, \bibinfo {author} {\bibfnamefont {H.~R.}\ \bibnamefont {Watts}},
  \bibinfo {author} {\bibfnamefont {D.~J.}\ \bibnamefont {Virley}}, \bibinfo
  {author} {\bibfnamefont {A.}~\bibnamefont {Lloyd}}, \bibinfo {author}
  {\bibfnamefont {E.~A.}\ \bibnamefont {Irving}}, \bibinfo {author}
  {\bibfnamefont {T.}~\bibnamefont {Nagafuji}}, \bibinfo {author}
  {\bibfnamefont {M.}~\bibnamefont {Ninomiya}}, \bibinfo {author}
  {\bibfnamefont {H.}~\bibnamefont {Nakamura}}, \bibinfo {author}
  {\bibfnamefont {A.~K.}\ \bibnamefont {Dunn}}, \ and\ \bibinfo {author}
  {\bibfnamefont {R.}~\bibnamefont {Graf}},\ }\bibfield  {title} {\enquote
  {\bibinfo {title} {{{P}eri-infarct depolarizations lead to loss of perfusion
  in ischaemic gyrencephalic cerebral cortex}},}\ }\href@noop {} {\bibfield
  {journal} {\bibinfo  {journal} {Brain}\ }\textbf {\bibinfo {volume} {130}},\
  \bibinfo {pages} {995--1008} (\bibinfo {year} {2007})}\BibitemShut {NoStop}%
\bibitem [{\citenamefont {Dahlem}\ and\ \citenamefont
  {M{\"u}ller}(1999)}]{DAH99}%
  \BibitemOpen
  \bibfield  {author} {\bibinfo {author} {\bibfnamefont {M.~A.}\ \bibnamefont
  {Dahlem}}\ and\ \bibinfo {author} {\bibfnamefont {S.~C.}\ \bibnamefont
  {M{\"u}ller}},\ }\bibfield  {title} {\enquote {\bibinfo {title} {Image
  processing techniques to analyse traveling waves},}\ }\href@noop {}
  {\bibfield  {journal} {\bibinfo  {journal} {Forma}\ }\textbf {\bibinfo
  {volume} {13}},\ \bibinfo {pages} {375--386} (\bibinfo {year}
  {1999})}\BibitemShut {NoStop}%
\bibitem [{\citenamefont {Dahlem}\ \emph {et~al.}(2010)\citenamefont {Dahlem},
  \citenamefont {Graf}, \citenamefont {Strong}, \citenamefont {Dreier},
  \citenamefont {Dahlem}, \citenamefont {Sieber}, \citenamefont {Hanke},
  \citenamefont {Podoll},\ and\ \citenamefont {Sch{\"o}ll}}]{DAH09a}%
  \BibitemOpen
  \bibfield  {author} {\bibinfo {author} {\bibfnamefont {M.~A.}\ \bibnamefont
  {Dahlem}}, \bibinfo {author} {\bibfnamefont {R.}~\bibnamefont {Graf}},
  \bibinfo {author} {\bibfnamefont {A.~J.}\ \bibnamefont {Strong}}, \bibinfo
  {author} {\bibfnamefont {J.~P.}\ \bibnamefont {Dreier}}, \bibinfo {author}
  {\bibfnamefont {Y.~A.}\ \bibnamefont {Dahlem}}, \bibinfo {author}
  {\bibfnamefont {M.}~\bibnamefont {Sieber}}, \bibinfo {author} {\bibfnamefont
  {W.}~\bibnamefont {Hanke}}, \bibinfo {author} {\bibfnamefont
  {K.}~\bibnamefont {Podoll}}, \ and\ \bibinfo {author} {\bibfnamefont
  {E.}~\bibnamefont {Sch{\"o}ll}},\ }\bibfield  {title} {\enquote {\bibinfo
  {title} {Two-dimensional wave patterns of spreading depolarization:
  retracting, re-entrant, and stationary waves},}\ }\href {\doibase
  doi:10.1016/j.physd.2009.08.009} {\bibfield  {journal} {\bibinfo  {journal}
  {Physica D}\ }\textbf {\bibinfo {volume} {239}},\ \bibinfo {pages} {889--903}
  (\bibinfo {year} {2010})}\BibitemShut {NoStop}%
\bibitem [{\citenamefont {Dreier}(2011)}]{DRE11}%
  \BibitemOpen
  \bibfield  {author} {\bibinfo {author} {\bibfnamefont {J.~P.}\ \bibnamefont
  {Dreier}},\ }\bibfield  {title} {\enquote {\bibinfo {title} {{{T}he role of
  spreading depression, spreading depolarization and spreading ischemia in
  neurological disease}},}\ }\href@noop {} {\bibfield  {journal} {\bibinfo
  {journal} {Nat. Med.}\ }\textbf {\bibinfo {volume} {17}},\ \bibinfo {pages}
  {439--447} (\bibinfo {year} {2011})}\BibitemShut {NoStop}%
\bibitem [{\citenamefont {Grafstein}(1963)}]{GRA63}%
  \BibitemOpen
  \bibfield  {author} {\bibinfo {author} {\bibfnamefont {B.}~\bibnamefont
  {Grafstein}},\ }\bibfield  {title} {\enquote {\bibinfo {title} {Neural
  release of potassium during spreading depression.}}\ }in\ \href@noop {}
  {\emph {\bibinfo {booktitle} {Brain Function. Cortical Excitability and
  Steady Potentials}}},\ \bibinfo {editor} {edited by\ \bibinfo {editor}
  {\bibfnamefont {M.~A.~B.}\ \bibnamefont {Brazier}}}\ (\bibinfo  {publisher}
  {University of California Press},\ \bibinfo {address} {Berkeley},\ \bibinfo
  {year} {1963})\ pp.\ \bibinfo {pages} {87--124}\BibitemShut {NoStop}%
\bibitem [{\citenamefont {Goadsby}, \citenamefont {Lipton},\ and\ \citenamefont
  {Ferrari}(2002)}]{GOA02}%
  \BibitemOpen
  \bibfield  {author} {\bibinfo {author} {\bibfnamefont {P.~J.}\ \bibnamefont
  {Goadsby}}, \bibinfo {author} {\bibfnamefont {R.~B.}\ \bibnamefont {Lipton}},
  \ and\ \bibinfo {author} {\bibfnamefont {M.~D.}\ \bibnamefont {Ferrari}},\
  }\bibfield  {title} {\enquote {\bibinfo {title} {{{M}igraine--current
  understanding and treatment}},}\ }\href@noop {} {\bibfield  {journal}
  {\bibinfo  {journal} {N.~Engl. J.~Med.}\ }\textbf {\bibinfo {volume} {346}},\
  \bibinfo {pages} {257--270} (\bibinfo {year} {2002})}\BibitemShut {NoStop}%
\bibitem [{\citenamefont {Schoenen}\ \emph
  {et~al.}(2013{\natexlab{a}})\citenamefont {Schoenen}, \citenamefont
  {Vandersmissen}, \citenamefont {Jeangette}, \citenamefont {Herroelen},
  \citenamefont {Vandenheede}, \citenamefont {G{\'e}rard},\ and\ \citenamefont
  {Magis}}]{SCH13d}%
  \BibitemOpen
  \bibfield  {author} {\bibinfo {author} {\bibfnamefont {J.}~\bibnamefont
  {Schoenen}}, \bibinfo {author} {\bibfnamefont {B.}~\bibnamefont
  {Vandersmissen}}, \bibinfo {author} {\bibfnamefont {S.}~\bibnamefont
  {Jeangette}}, \bibinfo {author} {\bibfnamefont {L.}~\bibnamefont
  {Herroelen}}, \bibinfo {author} {\bibfnamefont {M.}~\bibnamefont
  {Vandenheede}}, \bibinfo {author} {\bibfnamefont {P.}~\bibnamefont
  {G{\'e}rard}}, \ and\ \bibinfo {author} {\bibfnamefont {D.}~\bibnamefont
  {Magis}},\ }\bibfield  {title} {\enquote {\bibinfo {title} {Migraine
  prevention with a supraorbital transcutaneous stimulator. {A} randomized
  controlled trial},}\ }\href@noop {} {\bibfield  {journal} {\bibinfo
  {journal} {Neurology}\ }\textbf {\bibinfo {volume} {80}},\ \bibinfo {pages}
  {697--704} (\bibinfo {year} {2013}{\natexlab{a}})}\BibitemShut {NoStop}%
\bibitem [{\citenamefont {Lipton}\ \emph {et~al.}(2010)\citenamefont {Lipton},
  \citenamefont {Dodick}, \citenamefont {Silberstein}, \citenamefont {Saper},
  \citenamefont {Aurora}, \citenamefont {Pearlman}, \citenamefont {Fischell},
  \citenamefont {Ruppel},\ and\ \citenamefont {Goadsby}}]{LIP10}%
  \BibitemOpen
  \bibfield  {author} {\bibinfo {author} {\bibfnamefont {R.~B.}\ \bibnamefont
  {Lipton}}, \bibinfo {author} {\bibfnamefont {D.~W.}\ \bibnamefont {Dodick}},
  \bibinfo {author} {\bibfnamefont {S.~D.}\ \bibnamefont {Silberstein}},
  \bibinfo {author} {\bibfnamefont {J.~R.}\ \bibnamefont {Saper}}, \bibinfo
  {author} {\bibfnamefont {S.~K.}\ \bibnamefont {Aurora}}, \bibinfo {author}
  {\bibfnamefont {S.~H.}\ \bibnamefont {Pearlman}}, \bibinfo {author}
  {\bibfnamefont {R.~E.}\ \bibnamefont {Fischell}}, \bibinfo {author}
  {\bibfnamefont {P.~L.}\ \bibnamefont {Ruppel}}, \ and\ \bibinfo {author}
  {\bibfnamefont {P.~J.}\ \bibnamefont {Goadsby}},\ }\bibfield  {title}
  {\enquote {\bibinfo {title} {{S}ingle-pulse transcranial magnetic stimulation
  for acute treatment of migraine with aura: a randomised, double-blind,
  parallel-group, sham-controlled trial},}\ }\href@noop {} {\bibfield
  {journal} {\bibinfo  {journal} {Lancet Neurol.}\ }\textbf {\bibinfo {volume}
  {9}},\ \bibinfo {pages} {373--380} (\bibinfo {year} {2010})}\BibitemShut
  {NoStop}%
\bibitem [{\citenamefont {Ayata}(2010)}]{AYA10}%
  \BibitemOpen
  \bibfield  {author} {\bibinfo {author} {\bibfnamefont {C.}~\bibnamefont
  {Ayata}},\ }\bibfield  {title} {\enquote {\bibinfo {title} {Cortical
  spreading depression triggers migraine attack: Pro},}\ }\href {\doibase
  10.1111/j.1526-4610.2010.01647.x} {\bibfield  {journal} {\bibinfo  {journal}
  {Headache: The Journal of Head and Face Pain}\ }\textbf {\bibinfo {volume}
  {50}},\ \bibinfo {pages} {725--730} (\bibinfo {year} {2010})}\BibitemShut
  {NoStop}%
\bibitem [{\citenamefont {Dasilva}\ \emph {et~al.}(2012)\citenamefont
  {Dasilva}, \citenamefont {Mendonca}, \citenamefont {Zaghi}, \citenamefont
  {Lopes}, \citenamefont {Dossantos}, \citenamefont {Spierings}, \citenamefont
  {Bajwa}, \citenamefont {Datta}, \citenamefont {Bikson},\ and\ \citenamefont
  {Fregni}}]{DAS12}%
  \BibitemOpen
  \bibfield  {author} {\bibinfo {author} {\bibfnamefont {A.~F.}\ \bibnamefont
  {Dasilva}}, \bibinfo {author} {\bibfnamefont {M.~E.}\ \bibnamefont
  {Mendonca}}, \bibinfo {author} {\bibfnamefont {S.}~\bibnamefont {Zaghi}},
  \bibinfo {author} {\bibfnamefont {M.}~\bibnamefont {Lopes}}, \bibinfo
  {author} {\bibfnamefont {M.~F.}\ \bibnamefont {Dossantos}}, \bibinfo {author}
  {\bibfnamefont {E.~L.}\ \bibnamefont {Spierings}}, \bibinfo {author}
  {\bibfnamefont {Z.}~\bibnamefont {Bajwa}}, \bibinfo {author} {\bibfnamefont
  {A.}~\bibnamefont {Datta}}, \bibinfo {author} {\bibfnamefont
  {M.}~\bibnamefont {Bikson}}, \ and\ \bibinfo {author} {\bibfnamefont
  {F.}~\bibnamefont {Fregni}},\ }\bibfield  {title} {\enquote {\bibinfo {title}
  {{t{D}{C}{S}-induced analgesia and electrical fields in pain-related neural
  networks in chronic migraine}},}\ }\href@noop {} {\bibfield  {journal}
  {\bibinfo  {journal} {Headache}\ }\textbf {\bibinfo {volume} {52}},\ \bibinfo
  {pages} {1283--1295} (\bibinfo {year} {2012})}\BibitemShut {NoStop}%
\bibitem [{\citenamefont {Antal}\ \emph {et~al.}(2011)\citenamefont {Antal},
  \citenamefont {Kriener}, \citenamefont {Lang}, \citenamefont {Boros},\ and\
  \citenamefont {Paulus}}]{ANT11a}%
  \BibitemOpen
  \bibfield  {author} {\bibinfo {author} {\bibfnamefont {A.}~\bibnamefont
  {Antal}}, \bibinfo {author} {\bibfnamefont {N.}~\bibnamefont {Kriener}},
  \bibinfo {author} {\bibfnamefont {N.}~\bibnamefont {Lang}}, \bibinfo {author}
  {\bibfnamefont {K.}~\bibnamefont {Boros}}, \ and\ \bibinfo {author}
  {\bibfnamefont {W.}~\bibnamefont {Paulus}},\ }\bibfield  {title} {\enquote
  {\bibinfo {title} {{{C}athodal transcranial direct current stimulation of the
  visual cortex in the prophylactic treatment of migraine}},}\ }\href@noop {}
  {\bibfield  {journal} {\bibinfo  {journal} {Cephalalgia}\ }\textbf {\bibinfo
  {volume} {31}},\ \bibinfo {pages} {820--828} (\bibinfo {year}
  {2011})}\BibitemShut {NoStop}%
\bibitem [{\citenamefont {Burstein}\ and\ \citenamefont
  {Jakubowski}(2005)}]{BUR05a}%
  \BibitemOpen
  \bibfield  {author} {\bibinfo {author} {\bibfnamefont {R.}~\bibnamefont
  {Burstein}}\ and\ \bibinfo {author} {\bibfnamefont {M.}~\bibnamefont
  {Jakubowski}},\ }\bibfield  {title} {\enquote {\bibinfo {title} {Unitary
  hypothesis for multiple triggers of the pain and strain of migraine},}\
  }\href@noop {} {\bibfield  {journal} {\bibinfo  {journal} {J.~Comp.
  Neurology}\ }\textbf {\bibinfo {volume} {493}},\ \bibinfo {pages} {9--14}
  (\bibinfo {year} {2005})}\BibitemShut {NoStop}%
\bibitem [{\citenamefont {Goadsby}\ and\ \citenamefont
  {Silberstein}(2013)}]{GOA13a}%
  \BibitemOpen
  \bibfield  {author} {\bibinfo {author} {\bibfnamefont {P.~J.}\ \bibnamefont
  {Goadsby}}\ and\ \bibinfo {author} {\bibfnamefont {S.~D.}\ \bibnamefont
  {Silberstein}},\ }\bibfield  {title} {\enquote {\bibinfo {title} {{{M}igraine
  triggers: harnessing the messages of clinical practice}},}\ }\href@noop {}
  {\bibfield  {journal} {\bibinfo  {journal} {Neurology}\ }\textbf {\bibinfo
  {volume} {80}},\ \bibinfo {pages} {424--425} (\bibinfo {year}
  {2013})}\BibitemShut {NoStop}%
\bibitem [{\citenamefont {Hougaard}\ \emph {et~al.}(2013)\citenamefont
  {Hougaard}, \citenamefont {Amin}, \citenamefont {Amin}, \citenamefont
  {Hauge}, \citenamefont {Ashina},\ and\ \citenamefont {Olesen}}]{HOU13}%
  \BibitemOpen
  \bibfield  {author} {\bibinfo {author} {\bibfnamefont {A.}~\bibnamefont
  {Hougaard}}, \bibinfo {author} {\bibfnamefont {F.~M.}\ \bibnamefont {Amin}},
  \bibinfo {author} {\bibfnamefont {F.}~\bibnamefont {Amin}}, \bibinfo {author}
  {\bibfnamefont {A.~W.}\ \bibnamefont {Hauge}}, \bibinfo {author}
  {\bibfnamefont {M.}~\bibnamefont {Ashina}}, \ and\ \bibinfo {author}
  {\bibfnamefont {J.}~\bibnamefont {Olesen}},\ }\bibfield  {title} {\enquote
  {\bibinfo {title} {{{P}rovocation of migraine with aura using natural trigger
  factors}},}\ }\href@noop {} {\bibfield  {journal} {\bibinfo  {journal}
  {Neurology}\ }\textbf {\bibinfo {volume} {80}},\ \bibinfo {pages} {428--431}
  (\bibinfo {year} {2013})}\BibitemShut {NoStop}%
\bibitem [{\citenamefont {Paemeleire}\ and\ \citenamefont
  {Goodman}(2012)}]{PAE12}%
  \BibitemOpen
  \bibfield  {author} {\bibinfo {author} {\bibfnamefont {K.}~\bibnamefont
  {Paemeleire}}\ and\ \bibinfo {author} {\bibfnamefont {A.~M.}\ \bibnamefont
  {Goodman}},\ }\bibfield  {title} {\enquote {\bibinfo {title} {{{R}esults of a
  patient survey for an implantable neurostimulator to treat migraine
  headaches}},}\ }\href@noop {} {\bibfield  {journal} {\bibinfo  {journal}
  {J.~Headache Pain}\ }\textbf {\bibinfo {volume} {13}},\ \bibinfo {pages}
  {239--241} (\bibinfo {year} {2012})}\BibitemShut {NoStop}%
\bibitem [{\citenamefont {Tepper}\ \emph {et~al.}(2009)\citenamefont {Tepper},
  \citenamefont {Rezai}, \citenamefont {Narouze}, \citenamefont {Steiner},
  \citenamefont {Mohajer},\ and\ \citenamefont {Ansarinia}}]{TEP09}%
  \BibitemOpen
  \bibfield  {author} {\bibinfo {author} {\bibfnamefont {S.~J.}\ \bibnamefont
  {Tepper}}, \bibinfo {author} {\bibfnamefont {A.}~\bibnamefont {Rezai}},
  \bibinfo {author} {\bibfnamefont {S.}~\bibnamefont {Narouze}}, \bibinfo
  {author} {\bibfnamefont {C.}~\bibnamefont {Steiner}}, \bibinfo {author}
  {\bibfnamefont {P.}~\bibnamefont {Mohajer}}, \ and\ \bibinfo {author}
  {\bibfnamefont {M.}~\bibnamefont {Ansarinia}},\ }\bibfield  {title} {\enquote
  {\bibinfo {title} {Acute treatment of intractable migraine with
  sphenopalatine ganglion electrical stimulation},}\ }\href {\doibase
  10.1111/j.1526-4610.2009.01451.x} {\bibfield  {journal} {\bibinfo  {journal}
  {Headache}\ }\textbf {\bibinfo {volume} {49}},\ \bibinfo {pages} {983--989}
  (\bibinfo {year} {2009})}\BibitemShut {NoStop}%
\bibitem [{\citenamefont {Schoenen}\ \emph
  {et~al.}(2013{\natexlab{b}})\citenamefont {Schoenen}, \citenamefont {Jensen},
  \citenamefont {Lanteri-Minet}, \citenamefont {Lainez}, \citenamefont {Gaul},
  \citenamefont {Goodman}, \citenamefont {Caparso},\ and\ \citenamefont
  {May}}]{SCH13h}%
  \BibitemOpen
  \bibfield  {author} {\bibinfo {author} {\bibfnamefont {J.}~\bibnamefont
  {Schoenen}}, \bibinfo {author} {\bibfnamefont {R.~H.}\ \bibnamefont
  {Jensen}}, \bibinfo {author} {\bibfnamefont {M.}~\bibnamefont
  {Lanteri-Minet}}, \bibinfo {author} {\bibfnamefont {M.~J.}\ \bibnamefont
  {Lainez}}, \bibinfo {author} {\bibfnamefont {C.}~\bibnamefont {Gaul}},
  \bibinfo {author} {\bibfnamefont {A.~M.}\ \bibnamefont {Goodman}}, \bibinfo
  {author} {\bibfnamefont {A.}~\bibnamefont {Caparso}}, \ and\ \bibinfo
  {author} {\bibfnamefont {A.}~\bibnamefont {May}},\ }\bibfield  {title}
  {\enquote {\bibinfo {title} {{{S}timulation of the sphenopalatine ganglion
  ({S}{P}{G}) for cluster headache treatment. {P}athway {C}{H}-1: {A}
  randomized, sham-controlled study}},}\ }\href@noop {} {\bibfield  {journal}
  {\bibinfo  {journal} {Cephalalgia}\ }\textbf {\bibinfo {volume} {33}},\
  \bibinfo {pages} {816--830} (\bibinfo {year}
  {2013}{\natexlab{b}})}\BibitemShut {NoStop}%
\bibitem [{\citenamefont {Son}\ \emph {et~al.}(2012)\citenamefont {Son},
  \citenamefont {Cho}, \citenamefont {Kim}, \citenamefont {Choi}, \citenamefont
  {Lee}, \citenamefont {Chi}, \citenamefont {Park},\ and\ \citenamefont
  {Kim}}]{SON12}%
  \BibitemOpen
  \bibfield  {author} {\bibinfo {author} {\bibfnamefont {Y.-D.}\ \bibnamefont
  {Son}}, \bibinfo {author} {\bibfnamefont {Z.-H.}\ \bibnamefont {Cho}},
  \bibinfo {author} {\bibfnamefont {H.-K.}\ \bibnamefont {Kim}}, \bibinfo
  {author} {\bibfnamefont {E.-J.}\ \bibnamefont {Choi}}, \bibinfo {author}
  {\bibfnamefont {S.-Y.}\ \bibnamefont {Lee}}, \bibinfo {author} {\bibfnamefont
  {J.-G.}\ \bibnamefont {Chi}}, \bibinfo {author} {\bibfnamefont {C.-W.}\
  \bibnamefont {Park}}, \ and\ \bibinfo {author} {\bibfnamefont {Y.-B.}\
  \bibnamefont {Kim}},\ }\bibfield  {title} {\enquote {\bibinfo {title}
  {Glucose metabolism of the midline nuclei raphe in the brainstem observed by
  pet--mri fusion imaging},}\ }\href@noop {} {\bibfield  {journal} {\bibinfo
  {journal} {Neuroimage}\ }\textbf {\bibinfo {volume} {59}},\ \bibinfo {pages}
  {1094--1097} (\bibinfo {year} {2012})}\BibitemShut {NoStop}%
\bibitem [{\citenamefont {Cressman~Jr.}\ \emph {et~al.}(2009)\citenamefont
  {Cressman~Jr.}, \citenamefont {Ullah}, \citenamefont {Ziburkus},
  \citenamefont {Schiff},\ and\ \citenamefont {Barreto}}]{CRE08}%
  \BibitemOpen
  \bibfield  {author} {\bibinfo {author} {\bibfnamefont {J.~R.}\ \bibnamefont
  {Cressman~Jr.}}, \bibinfo {author} {\bibfnamefont {G.}~\bibnamefont {Ullah}},
  \bibinfo {author} {\bibfnamefont {J.}~\bibnamefont {Ziburkus}}, \bibinfo
  {author} {\bibfnamefont {S.~J.}\ \bibnamefont {Schiff}}, \ and\ \bibinfo
  {author} {\bibfnamefont {E.}~\bibnamefont {Barreto}},\ }\bibfield  {title}
  {\enquote {\bibinfo {title} {The influence of sodium and potassium dynamics
  on excitability, seizures, and the stability of persistent states: I. single
  neuron dynamics},}\ }\href {\doibase 10.1007/s10827-008-0132-4} {\bibfield
  {journal} {\bibinfo  {journal} {J.~Comput. Neurosci.}\ }\textbf {\bibinfo
  {volume} {26}},\ \bibinfo {pages} {159--170} (\bibinfo {year}
  {2009})}\BibitemShut {NoStop}%
\end{thebibliography}

%

\end{document}